\documentclass[12pt]{article}

\usepackage[a4paper,left=30mm,right=20mm,top=20mm,bottom=20mm]{geometry}
\usepackage{graphics}
\usepackage{amsmath}
\usepackage{epsfig}
\usepackage{cite}
\usepackage{xcolor}

\title{Associated Higgs + jet(s) production at the LHC and CCFM gluon dynamics in a proton}

\author{A.V.~Lipatov$^{1,2}$\footnote{e-mail: lipatov@theory.sinp.msu.ru}, M.A.~Malyshev$^{1}$ }

\begin{document}

\maketitle

\begin{center}

{\it $^{1}$Skobeltsyn Institute of Nuclear Physics, Lomonosov Moscow State University, 119991, Moscow, Russia}\\
{\it $^{2}$Joint Institute for Nuclear Research, 141980, Dubna, Moscow region, Russia}\\

\end{center}

\vspace{0.5cm}

\begin{center}

{\bf Abstract }

\end{center}

\indent

We consider the associated production of Higgs boson and hadronic jet(s) in $pp$ collisions at the LHC 
for the first time using the $k_T$-factorization
approach. Our analysis is based on the off-shell gluon-gluon fusion subprocess, where
non-zero transverse momenta of initial gluons are taken into account and
covers $H \to \gamma \gamma$, $H \to ZZ^* \to 4l$ (with $l = e,\mu$) and
$H \to W^+ W^- \to e^\pm \mu^\mp \nu \bar \nu$ decay channels.
The transverse momentum dependent (TMD) gluon densities in a proton are taken from
Catani-Ciafaloni-Fiorani-Marchesini evolution equation.
To simulate the kinematics of the produced jets the TMD parton shower
implemented into the Monte-Carlo event generator \textsc{cascade} is applied.
The comparison of our results with the latest  experimental data taken by the CMS and
ATLAS Collaborations at $\sqrt s = 8$ and $13$~TeV and conventional higher-order perturbative QCD
calculations is presented.
We highlight observables, which are sensitive to the TMD gluon densities in a proton.

\vspace{1.0cm}


\newpage

Recently, the ATLAS and CMS Collaborations have presented measurements\cite{1,2,3,4,5,6,7,8} of the
total and differential cross sections of Higgs boson production in $pp$ collisions at the LHC conditions,
both inclusive and associated with one or more hadronic jets.
These data have been taken for $H \to \gamma \gamma$, $H \to ZZ^* \to 4l$ (where $l = e$ or $\mu$)
and $H \to W^+W^- \to e^\pm \mu^\mp \nu \bar \nu$ decay channels at the center-of-mass
energies $\sqrt s = 8$ and $13$~TeV. Such measurements allow to probe fundamental properties of Higgs boson
(for example, spin and couplings to gauge bosons and fermions) and 
provide a testing ground for perturbative Quantum Chromodynamics
(pQCD) predictions. Moreover, they can be used to investigate the gluon dynamics in a proton since
the dominant mechanism of inclusive Higgs production is the gluon-gluon fusion (see, for example,\cite{9} and 
references therein).

The reported measurements\cite{1,2,3,4,5,6,7,8} are found to be in good agreement with the
next-to-next-to-leading-order (NNLO) pQCD
predictions\cite{10,11,12,13,14,15} obtained using \textsc{hres}\cite{16} and/or \textsc{nnlops}\cite{17} 
Monte-Carlo tools\footnote{The N$^3$LO predictions for gluon-gluon fusion subprocess have become available recently\cite{18}
and the NLO perturbative electroweak corrections to the Higgs production have been evaluated\cite{19,20,21,22,23}.}.
The available NNLO calculations
can be even improved 
at low transverse momenta 
by the soft-gluon resummation procedure,
which has been carried out up to next-to-next-to-leading logarithmic accuracy (NNLL)\cite{24,25}.
An alternative description of the LHC data\cite{1,2,3,4,5,6,7,8} can
be achieved in the framework of the high-energy QCD factorization\cite{26},
or $k_T$-factorization approach of QCD\cite{27}.
This approach is based on the Balitsky-Fadin-Kuraev-Lipatov (BFKL)\cite{28} or
Catani-Ciafaloni-Fiorani-Marchesini (CCFM)\cite{29} gluon evolution
equations, which resum large logarithmic terms proportional to $\alpha_s^n \ln^n s \sim \alpha_s^n \ln^n 1/x$,
important at high energies $\sqrt s$ (or, equivalently, at small longitudinal momentum fraction
$x$ of the colliding proton carried by an interacting gluon).
It can be understood since
typical $x$ values are about of $x \sim m_H/\sqrt s \sim 0.008 - 0.015$ for Higgs mass $m_H = 125$~GeV
and $\sqrt s = 13$~TeV.
Additionally, the CCFM equation takes into account terms proportional to $\alpha_s^n \ln^n 1/(1 - x)$
and therefore can be applied for both small and large $x$\cite{29}.
The $k_T$-factorization QCD approach has certain technical advantages in the ease of
including higher-order pQCD radiative corrections (namely, main part of NLO + NNLO + ... terms
corresponding to real initial-state gluon emissions) in the form of
transverse momentum dependent (TMD, or unintegrated) gluon density
and can be used as a convenient alternative to explicit higher-order pQCD calculations. 
The detailed description of this approach can be found, for example, in reviews\cite{30,31}.

The $k_T$-factorization formalism was applied\cite{32,33,34,35,36,37,38,39,40,41} to the inclusive Higgs boson
production at the LHC. As it was demostrated\cite{38,39,40,41},
this approach, being supplemented with the CCFM gluon dynamics, is able to
describe the data obtained at $\sqrt s = 8$ and $13$~TeV
in the diphoton, four-lepton and $H \to W^+W^- \to e^\pm \mu^\mp \nu \bar \nu$ decay
channels even with leading-order off-shell (depending on the non-zero transverse momenta of incoming gluons)
production amplitudes.
Comparison with the higher-order pQCD calculations
was presented\cite{41}.
The sensitivity of the $k_T$-factorization predictions to the
TMD gluon density specially was pointed out\cite{41,42,43}.


The associated production of Higgs boson and one or more hadronic jets 
is of special interest from different points of view.
In our opinion, the most intriguing and remarkable point is 
connected with the distinctive feature of the $k_T$-factorization approach 
regarding the final state jet formation.
While in the conventional (DGLAP-based) parton level pQCD the produced jets
are fully determined by corresponding hard scattering amplitude,
in the $k_T$-factorization scenario in addition to the quarks and/or gluons
produced in the hard subprocesses (which can form the hadronic jets)
there is a number of gluons radiated in the course of their non-collinear 
evolution, which also give rise to final state jets.
So, the measured events with the detected jets could be useful in
discrimination between the two calculation schemes. 
Therefore, it is of interest and importance to generate
the $k_T$-factorization predictions for such events
and, of course, test these predictions in as many cases as possible. 
Closely related to this is selection of the TMD gluon 
densities in a proton best suited to describe the available 
experimental data. 

In the present note we extend the previous 
consideration\cite{33,38,41} of inclusive Higgs production  
to $H + $ jet(s) events. 
Such calculations are performed for the first time.
To correctly implement into our evaluations the kinematics of the final state jets, 
the method of\cite{44} is applied.
This method is based on the reconstruction of 
CCFM evolution ladder using a TMD parton shower routine 
implemented into the Monte-Carlo event generator \textsc{cascade}\cite{45}. 

Our main formulas were obtained in previous papers\cite{33,38,41}.
However, for the reader's convenience, let us very shortly describe the basic calculation steps.
We start from the off-shell gluon fusion
subprocesses:
\begin{gather}
  g^*(k_1) + g^*(k_2) \to H(p) \to V(p_1) + V(p_2),
\end{gather}
\noindent
where four-momenta of all particles are indicated in the parentheses and 
$V$ denotes $\gamma$, $W^\pm$ or $Z$ bosons (any of the gauge bosons can decay into leptons and/or neutrino).
It is important that both initial gluons carry non-zero transverse momenta: ${\mathbf k}_{1T}^2 = - k_{1T}^2 \neq 0$, 
${\mathbf k}_{2T}^2 = - k_{2T}^2 \neq 0$.
Using the effective Lagrangian for the Higgs coupling to gluons\cite{46,47} valid in the limit of
infinite top quark mass, $m_t \to \infty$, one can easily obtain the corresponding 
off-shell  
production amplitudes. The latter can be written in a form\cite{38,41} (see also\cite{39}):
\begin{gather} 
  |{\cal \bar A}|^2 = {1\over 1152 \pi^4} \alpha^2 \alpha_s^2 G_F^2 |{\cal F}|^2 {\hat s^2 (\hat s + {\mathbf p}_T^2)^2 \over (\hat s - m_H^2)^2 + m_H^2 \Gamma_H^2} \cos^2 \phi,
  \label{eqHiggsAmpl1}
\end{gather}
\noindent
for $H \to \gamma \gamma$ decay and
\begin{gather} 
  |{\cal \bar A}|^2 = {512\pi \over 9} \alpha^3 \alpha_s^2 G_F \sqrt 2 m_Z^2 C_V { (\hat s + {\mathbf p}_T^2)^2 \over (\hat s - m_H^2)^2 + m_H^2 \Gamma_H^2} \cos^2 \phi \times \nonumber \\
  \times {(g_{(V)L}^4 + g_{(V)R}^4) (l_1 \cdot l_3) (l_2 \cdot l_4) + 2 g_{(V)L}^2 \, g_{(V)R}^2 (l_1 \cdot l_4) (l_2 \cdot l_3) \over [(p_1^2 - m_V^2)^2 + m_V^2\Gamma_V^2] [(p_2^2 - m_V^2)^2 + m_V^2\Gamma_V^2]},
  \label{eqHiggsAmpl2}
\end{gather}
\noindent
for $H \to ZZ^* \to 4l$ or $H \to W^+ W^- \to e^\pm \mu^\mp \nu \bar \nu$ decays.
Here $G_F$ is the Fermi coupling constant, $l_1$ and $l_3$ are the gauge bosons decay leptons four-momenta,
$l_2$ and $l_4$ are their antileptons four-momenta (so that $p_1 = l_1 + l_2$ and 
$p_2 = l_3 + l_4$), $\hat s = (k_1 + k_2)^2$, ${\mathbf p}_T = {\mathbf k}_{1T} + {\mathbf k}_{2T}$ is 
the transverse momentum of the Higgs boson, $\phi$ is the azimuthal angle between 
the transverse momenta of initial off-shell gluons, $m_V$ and $\Gamma_V$ are the masses and 
decay widths of corresponding particles.
The exact expressions for ${\cal F}$, 
$C_V$, left and right weak current constants $g_{(V)L}$ and $g_{(V)R}$ 
are listed\cite{38,41}, there all the calculation details are given. 
The gauge inavriant off-shell production amplitudes~(2) and (3) have been implemented into
the parton-level Monte-Carlo event generator \textsc{pegasus}\cite{48}.

An important point of our calculations is connected with the proper determination
of associated jets four-momenta. As it was noted above, 
the produced Higgs boson is accompanied by a number of gluons radiated in the course of the 
non-collinear evolution, which give rise to final jets.
Similar to\cite{44},
to reconsruct one or few leading hadronic jets
from all of these initial state gluon emissions
we have used the anti-$k_T$ algorithm with radia $R_{\rm jet}$
as implemented into the \textsc{fastjet} tool\cite{52}.
Technically, we generate a Les Houches Event file\cite{53}
in the \textsc{pegasus} calculations
and then process the file with a TMD shower tool implemented
into the Monte-Carlo event generator \textsc{cascade}\cite{45}.
In this way we reconstruct the CCFM evolution ladder
and consistently compute the cross section of associated $H$ + jet(s) production
according to the experimental setup\footnote{A simplified model to implement the effects 
of parton showers into analytical calculations was used in early calculations\cite{33}.}.
The CCFM equation seems to be the most suitable tool for our consideration because it
smoothly interpolates between the small-$x$ BFKL gluon dynamics and 
conventional DGLAP one.

Concerning the CCFM-evolved gluon densities in a proton,
in the present note we tested two different sets\footnote{A comprehensive collection of the TMD gluon
densities can be found in the \textsc{tmdlib} package\cite{51}, which is a C++ library 
providing a framework and interface to the different parametrizations.}, 
namely, JH'2013 set 2\cite{49} and (more old) A0 set\cite{50}.
The input parameters of latest gluon density, JH'2013 set 2,
have been derived from the best description of high-precision HERA data on proton structure functions
$F_2(x,Q^2)$ and $F_2^c(x,Q^2)$\cite{49}.

Throughout this paper, all the calculations are based on the following parameter setting.
We kept $n_f=4$ active (massless) quark flavours, set 
$\Lambda_{\rm QCD} = 200$($250$)~MeV and
used two-loop (one-loop) QCD coupling for JH'2013 set 2 (A0) gluon densities. 
As it is often done, the renormalization scale was taken to be $\mu_R^2 = m_H^2$. 
The factorization scale was taken as $\mu_F^2 = \hat s + {\mathbf Q}_T^2$
(where ${\mathbf Q}_T$ is the net transverse momentum of the initial
off-shell gluon pair),
that is dictated mainly by the CCFM evolution algorithm (see\cite{49,50} for more information).
To evaluate the theoretical uncertainties we use auxiliary gluon densities JH'2013 set 2$+$ 
and JH'2013 set 2$-$ as well as A0$+$ and A0$-$ instead of default gluon distribution functions.
These two sets refer to the varied hard scales in the strong coupling constant $\alpha_s$ in the off-shell amplitude: '$+$' 
stands for $2\mu_R$, while '$-$' refers to $\mu_R/2$. 
Following\cite{54}, we set electroweak and Higgs bosons masses
$m_Z = 91.1876$~GeV, $m_W = 80.403$~GeV and $m_H = 125.1$~GeV, their
total decay widths $\Gamma_Z = 2.4952$~GeV, $\Gamma_W = 2.085$~GeV and $\Gamma_H = 4.3$~MeV
and use $\sin^2\theta_W = 0.23122$.

We start the discussion by presenting our results
for associated Higgs boson and jet production in the diphoton decay channel.
The latest measurements were done by the CMS\cite{1} and ATLAS\cite{5} 
Collaborations at the $\sqrt s = 13$~TeV. The 
applied experimental cuts are collected in Table~1. 
An additional requirement (the isolation criterion) is introduced for the photons in both experiments: 
the sum of transverse energy of particles 
around every photon within the radius $\Delta R = \sqrt{\Delta \eta^2 + \Delta \phi^2} = 0.3$ has to be 
smaller than $E_{\rm iso} = 10$~GeV.
The results of our calculations are presented on Figs.~1 and 2. 
Note that here we concentrated only on some of the kinematical variables among the quite
large variety of those, presented by the CMS and ATLAS Collaborations: 
number of jets $N_{\rm jet}$, leading jet transverse momentum $p_T^{j_1}$ and rapidity $y^{j_1}$, 
rapidity difference between the diphoton system and  
leading jet, $\Delta y^{\gamma\gamma j_1}$, 
azimuthal angle difference between the two leading jets, $\Delta \phi^{j_1j_2}$
and difference between the 
average pseudorapidity of these jets and pseudorapidity of 
the diphoton system $|\eta^{j_1j_2} - \eta^{\gamma\gamma}|$ (Zeppenfeld variable\cite{55}).
We added the contributions from weak boson 
fusion subprocesses ($W^+W^- \to H$ and $ZZ \to H$), associated $HZ$ or $HW^\pm$
production and associated $t\bar t H$ production
to our results.
These contributions are essential at high transverse momenta 
and have been calculated in the conventional pQCD approach with the 
NLO accuracy (we took them from the CMS\cite{1} and ATLAS\cite{5} papers).
Also we show for comparison the NNLO pQCD predictions, calculated with the 
\textsc{nnlops} program\cite{17} and taken from\cite{1,5}.
As one can see, the measured cross sections can be, in general, reasonably well 
described by the $k_T$-factorization evaluations 
within the theoretical and experimental uncertainties.
However, the predictions based on the JH'2013 set 2 and A0 gluon densities behave
differently for some observables, especially for correlation observables,
such as $\Delta y^{\gamma\gamma j_1}$ and 
azimuthal angle difference $\Delta\phi^{\gamma\gamma j_1}$ (see Fig.~2). 
Moreover, "old" A0 one tends to underestimate the data although
giving relatively larger number of jets at 
larger transverse momenta.
Unfortunately, the current level of experimental accuracy does not allow us to favor one or 
another TMD gluon density in a proton.
More precise future measurements of such 
observables could be promising to distinguish between the latter.
The NNLO pQCD predictions
behave similarly to the A0 results 
(except for distribution in $\Delta y^{\gamma\gamma j_1}$), though having larger normalization.
The difference between the NNLO pQCD and JH'2013 set 2 results is clearly seen for 
$\Delta y^{\gamma\gamma j_1}$ and, in some sence, for $\Delta\phi^{\gamma\gamma j_1}$ observable.
Nevertheless, our calculations demonstrate the 
possibility of $k_T$-factorization approach 
supplemented with the CCFM gluon dynamics
to reasonably describe the collider data on events containing hadronic jets
in final state.

Next, we turn to the $H \to ZZ^* \to 4l$ and $H \to W^+W^- \to e^\pm \mu^\mp \nu \bar \nu$
decay modes.
In these channels, the available experimental data  
have been obtained by the CMS\cite{2,3,4} and ATLAS\cite{6,7,8} Collaborations
at $\sqrt s = 8$ and $13$~TeV.
The applied experimental cuts are listed in Tables~2 and~3, respectively.
Our results for several interesting observables,
namely, $N_{\rm jet}$, leading jet transverse momentum
$p_T^{j_1}$, rapidity difference between the Higgs boson and  
leading jet, $\Delta y^{H j_1}$, pseudorapidity and azimuthal angle difference between the leading and subleading jets, 
$\Delta \eta^{j_1j_2}$ and $\Delta \phi^{j_1j_2}$,
invariant masses of Higgs-leading jet system $m^{Hj_1}$ and Higgs-dijet system $m^{Hj_1j_2}$, 
are shown in Figs.~3 --- 6.
As in the case of diphoton decay mode, the contributions from weak boson 
fusion subprocess, associated $HZ$, $HW^\pm$ and $t\bar t H$ production
calculated in the NLO pQCD approximation and
taken from\cite{2,3,4,6,7,8} were added to the off-shell gluon-gluon fusion.
We find again that latest JH'2013 set 2 gluon distribution reasonably well describes the LHC data 
within the estimated theoretical uncertainties
whereas A0 gluon lacks normalization.
The measured data\cite{2,3,4,6,7,8} point on the following distinctive observables, 
which reveal the difference between JH'2013 set 2 and A0 gluon densities:
Higgs-jet rapidity difference $\Delta y^{Hj_1}$, difference in pseudorapidity between 
the leading and subleading jets $\Delta \eta^{j_1j_2}$
and invariant masses $m^{Hj_1}$ and $m^{Hj_1j_2}$. 
The latter demonstrate much larger cross section at relatively low invariant masses for A0 predictions, 
which is, in fact, in better 
agreement with the ATLAS data (see Fig.~5).
Thus, in more precise forthcoming experiments 
the highlighted observables,
in addition to the variables for inclusive Higgs production pointed earlier\cite{33,38,41,42,43},
could be promising to 
distinguish between the different TMD gluon densities
or to better constrain their parameters. 
Like as for $\gamma\gamma$ decay channel, we plot also the collinear NNLO pQCD 
results, taken from\cite{2,3,4,6,7,8}. It can be seen again, that the A0 distributions 
generally follow the collinear results in shape, whereas the JH'2013 set 2 
results somewhat differ from them.

To conclude, we calculated for the first time the cross sections of 
associated Higgs and jet(s) production at the LHC conditions 
using the $k_T$-factorization approach.
Our consideration covers different Higgs decay channels
and is mainly based on the off-shell production
amplitudes for gluon-gluon fusion subprocess (implemented into the Monte-Carlo event 
generator \textsc{pegasus}) and CCFM-evolved TMD gluon densities in a proton.
To reconstruct correctly the kinematics of the final-state hadronic jets the TMD
parton shower generator \textsc{cascade} has been applied.
Our predictions obtained with the recent JH'2013 set 2 gluon density
agree well with the experimental data taken by the CMS and ATLAS Collaborations
at $\sqrt s = 8$ and $13$~TeV.
We have found observables, which are sensitive to the TMD
gluon densities in a proton. 
As it was expected, these are the ones related with the 
properties of the produced jets, for example, the 
rapidity difference between the Higgs boson and leading jet.
Unfortunately, the current level of experimental accuracy does not allow to 
distinguish between the latter. However,
more precise future experimental studies 
of the pointed observables could be promising and could allow one to 
constrain the TMD gluons.
Our study demonstrates the possibility of 
$k_T$-factorization approach supplemented with the CCFM gluon dynamics 
to describe the 
events with large number of jets in final state. It
significantly extends the previous consideration\cite{33,38,41}
of inclusive Higgs production at the LHC.

{\it Acknowledgements.} We thank S.P.~Baranov and H.~Jung for their interest, very useful discussions on the topic
and important remarks.  We are grateful the DESY Directorate for the support in the framework of 
Cooperation Agreement between MSU and DESY on phenomenology of the LHC processes and TMD parton densities.
M.A.M. was also supported by the grant of the Foundation for the Advancement of Theoretical Physics and Mathematics “BASIS” 20-1-3-11-1.

\begin{table} \footnotesize 
\label{table1}
\begin{center}
\begin{tabular}{|c|c|c|c|c|}
\hline
 & ATLAS~\cite{5}   & CMS~\cite{1}  \\\hline
$p_T^{\gamma_1}$, GeV & $>0.35 m^{\gamma \gamma}$ & $>m^{\gamma \gamma}/3$  \\\hline
$p_T^{\gamma_2}$, GeV & $>0.25 m^{\gamma \gamma}$ & $>m^{\gamma \gamma}/4$  \\\hline
$|y^{\gamma}|$ & $<2.37$, & $<2.5$   \\\hline
$m^{\gamma \gamma}$, GeV & $105 - 160$ & $ > 90$    \\\hline
$R_{\text {jet}}$ & \multicolumn{2}{c|}{0.4}    \\\hline
$p_T^{\text {jet}}$, GeV & $>30$  & $>30$  \\\hline
$|y^\text{jet}|$ & $<4.4$ & $<4.7$   \\\hline
\end{tabular}
\end{center}
\caption{Basic parameters, used for simulations in the $H \to \gamma \gamma$ decay channel.}
\end{table}

\begin{table} \footnotesize 
\label{table2}
\begin{center}
\begin{tabular}{|c|c|c|c|c|}
\hline
 & ATLAS~\cite{8}   & CMS~\cite{4}  \\\hline
$p_T^{l_1}$, GeV & $>22$ & $>25$   \\\hline
$p_T^{l_2}$, GeV & $>15$ & $>13$    \\\hline
$|\eta^l|$ & $<2.47$, excl. $1.37<|\eta^l|<1.52$ $(<2.5)$ & $<2.5$   \\\hline
$m^{ll}$, GeV & $10-55$ & $>12$    \\\hline
$R_{\text {jet}}$ & \multicolumn{2}{c|}{0.4}    \\\hline
$p_T^{\text {jet}}$, GeV & $>25$, if $|\eta^\text{jet}|<2.4$, $>30$ otherwise  & $>30$  \\\hline
$|\eta^\text{jet}|$ & $<4.5$ & $<4.7$   \\\hline
other cuts & $p_T^\text{miss}>20$ GeV & $p_T^{ll}>30$ GeV  \\
& $\Delta\phi^{ll}<1.8$ & $m_T^{l_2}\equiv\sqrt{2p_T^{l_2}p_T^\text{miss}[1-\cos\Delta\phi(\mathbf p_T^{l_2},\mathbf p_T^\text{miss})]}>30$~GeV  \\
&  & $m_T^H\equiv\sqrt{2p_T^{ll}p_T^\text{miss}[1-\cos\Delta\phi(\mathbf p_T^{ll},\mathbf p_T^\text{miss})]}>60$~GeV   \\\hline
\end{tabular}
\end{center}
\caption{Basic parameters, used for simulations in the $H \to ZZ^* \to 4l$ decay channel. By default experimental 
cuts for electrons are shown. Cuts for muons are placed in brackets, if differ.}
\end{table}

\begin{table} \footnotesize 
\label{table3}
\begin{center}
\begin{tabular}{|c|c|c|c|c|}
\hline
 & CMS   & ATLAS & CMS & ATLAS \\
 & {\footnotesize $\sqrt{s}=8$~TeV \cite{2}}  & {\footnotesize $\sqrt{s}=8$~TeV \cite{6}}  & {\footnotesize $\sqrt{s}=13$~TeV \cite{3}} & {\footnotesize $\sqrt{s}=13$~TeV \cite{7}} \\\hline
ordered $p_T^l$, GeV & $>20, 10, 7(5), 7(5)$ & $>20, 15, 10, 7(6)$ & $>20, 10, 7(5), 7(5)$ & $>20, 15, 10, 5$  \\\hline
$|\eta^l|$ & $<2.5 (2.4)$ & $<2.7 (2.47)$  & $<2.5 (2.4)$ & $<2.7$  \\\hline
$m^Z$ & $40-120$ & $50-106$ & $40-120$ & $50-106$  \\\hline
$m^{Z^*}$, GeV & $12-120$ & $12-115$ & $12-120$ & $12-115$   \\\hline
$m^{4l}$, GeV & $105-140$ & $118-129$ & $105-140$ & $105-160$   \\\hline
$\Delta R^\text{same-sign leptons}$ & $>0.02$ & $>0.1$ & $>0.02$  & $>0.1$ \\\hline
$\Delta R^\text{opposite-sign leptons}$ & $>0.02$ & $>0.2$ & $>0.02$ & $>0.1$  \\\hline
lepton isolation & $\Delta R=0.4$ &  & $\Delta R=0.3$ &  \\
 & $E_\text{iso}=0$ &   & $E_\text{iso}=0.35p_T^l$ &   \\\hline
$R_{\text {jet}}$ & 0.5 & 0.4 & 0.4 & 0.4 \\\hline
$p_T^{\text {jet}}$, GeV & $>30$ & $>30$ & $>30$ & $>30$ \\\hline
$\eta^{\text {jet}}$ & $<4.7$ & $<4.4$ & $<2.5$ & $<4.4$ \\\hline
\end{tabular}
\end{center}
\caption{Basic parameters, used for simulations in the $H \to W^+W^- \to e^\pm \mu^\mp \nu \bar \nu$ decay channel. By default experimental 
cuts for electrons are shown. Cuts for muons are placed in brackets, if differ.}
\end{table}

\newpage

\begin{figure}
\begin{center}
\includegraphics[width=7.9cm]{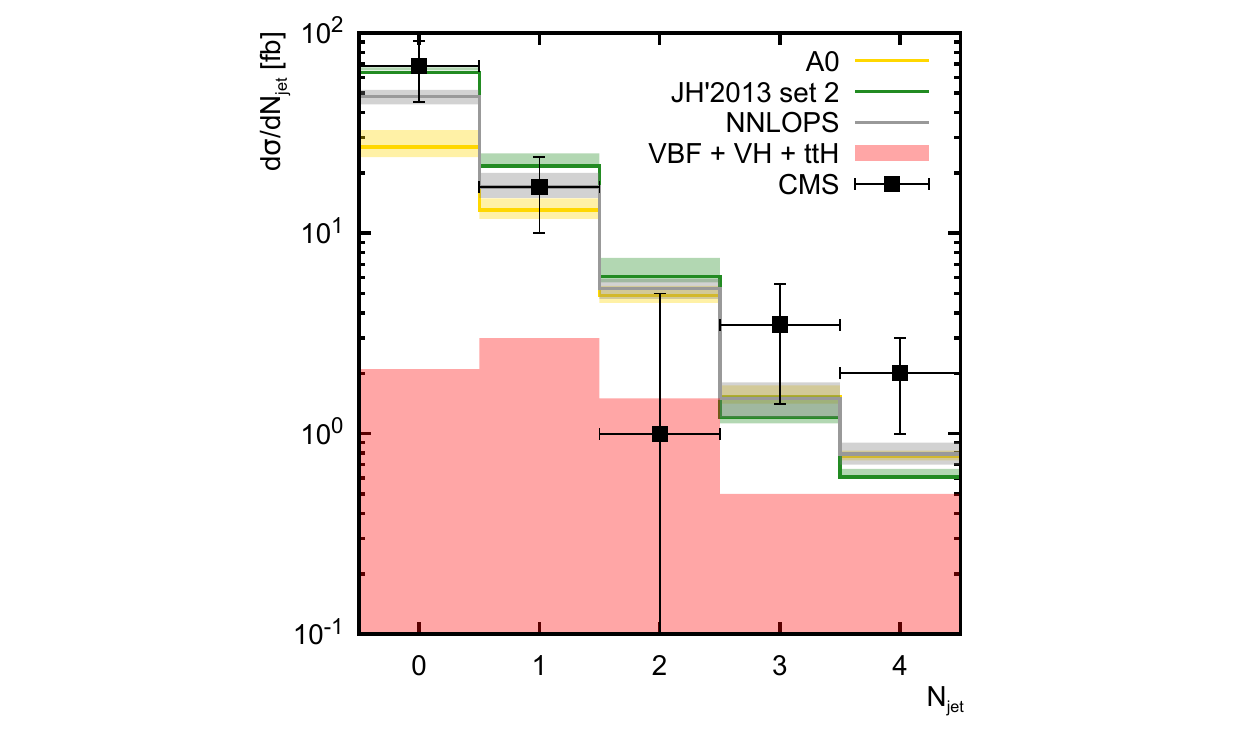}
\includegraphics[width=7.9cm]{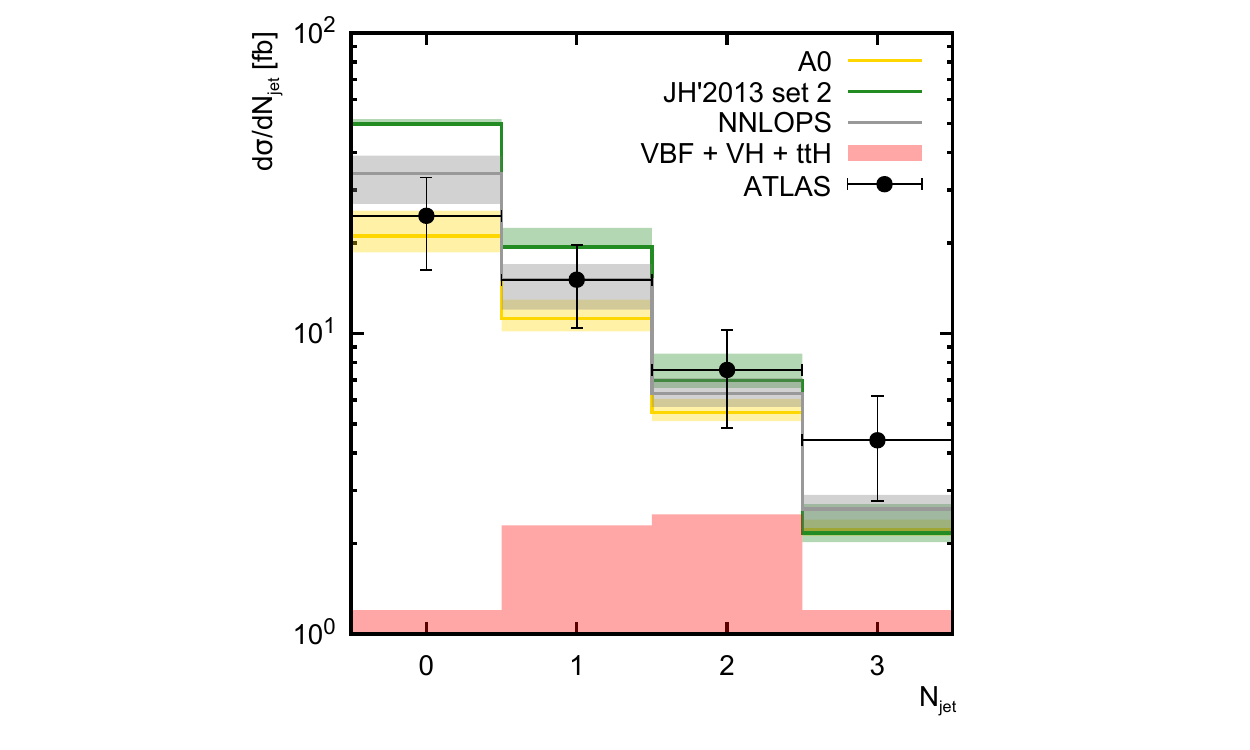}
\includegraphics[width=7.9cm]{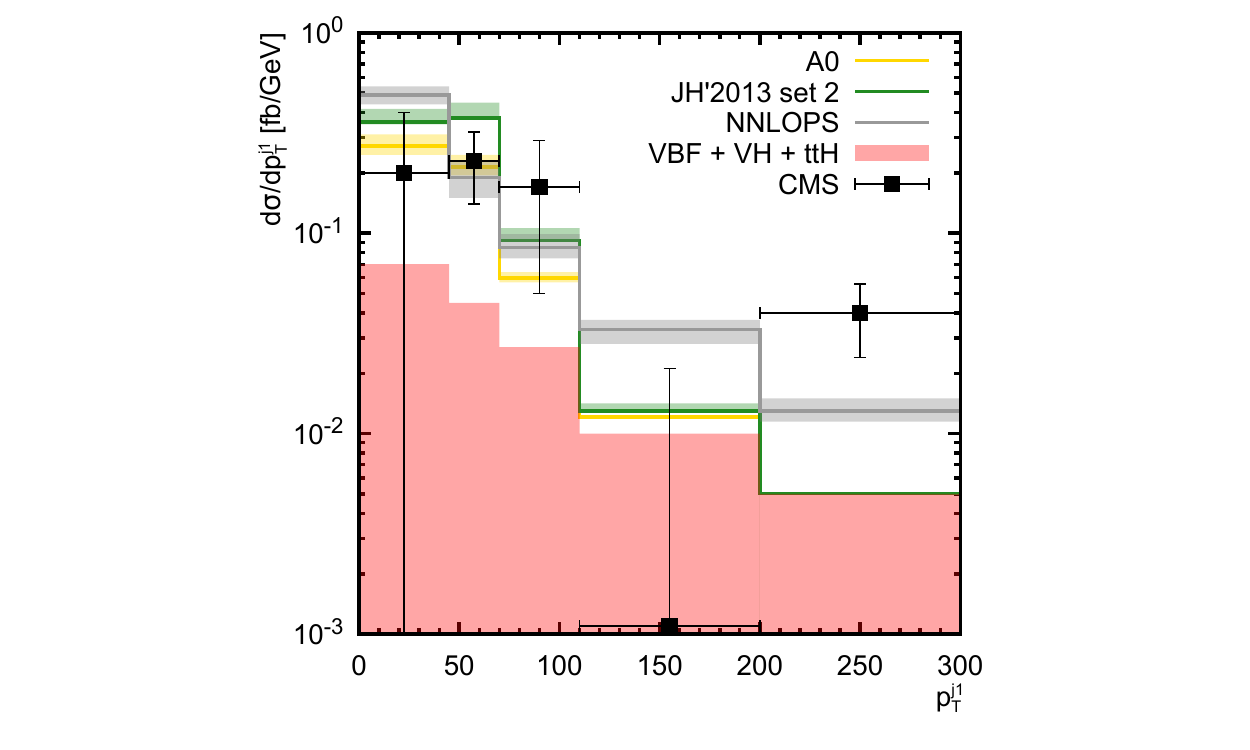}
\includegraphics[width=7.9cm]{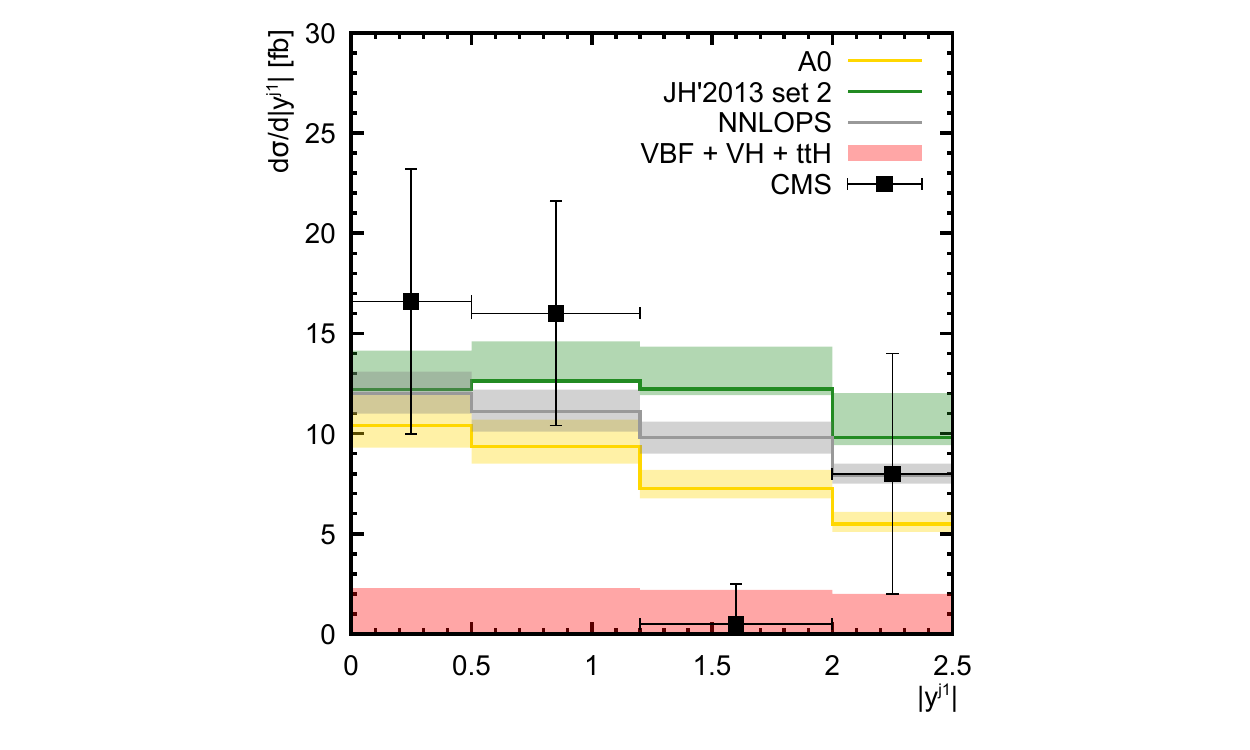}
\includegraphics[width=7.9cm]{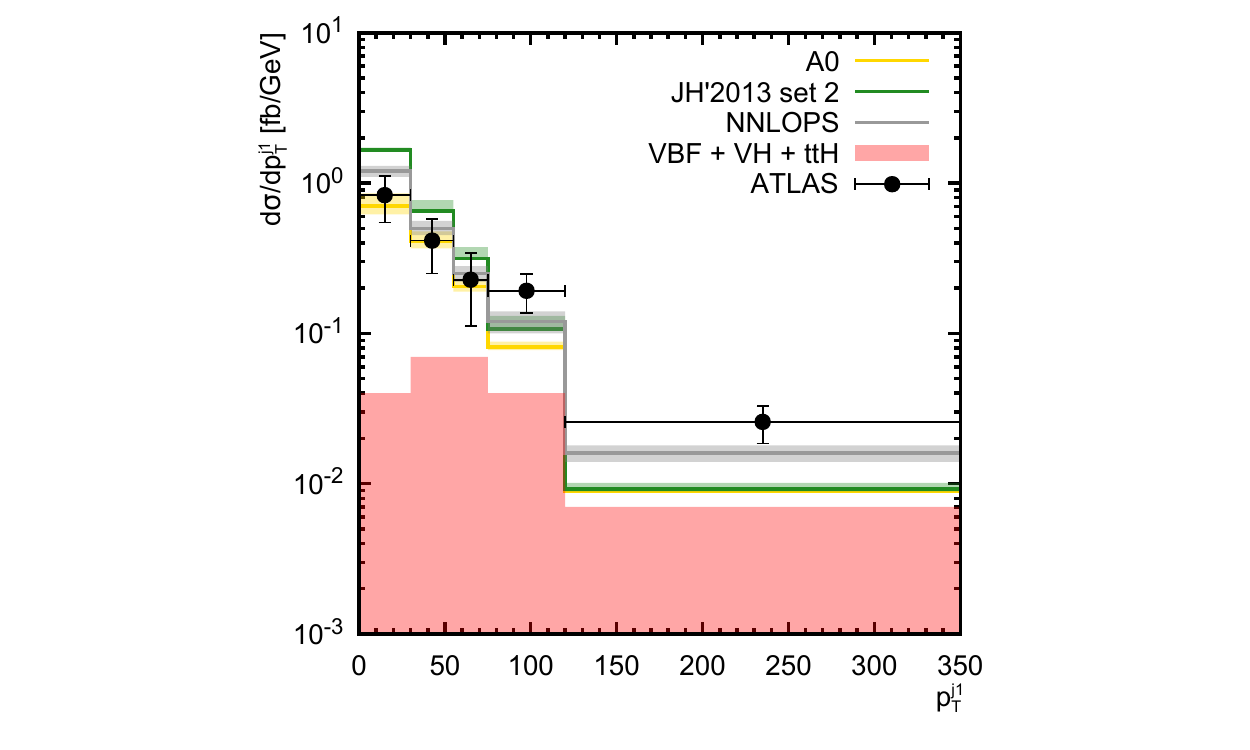}
\includegraphics[width=7.9cm]{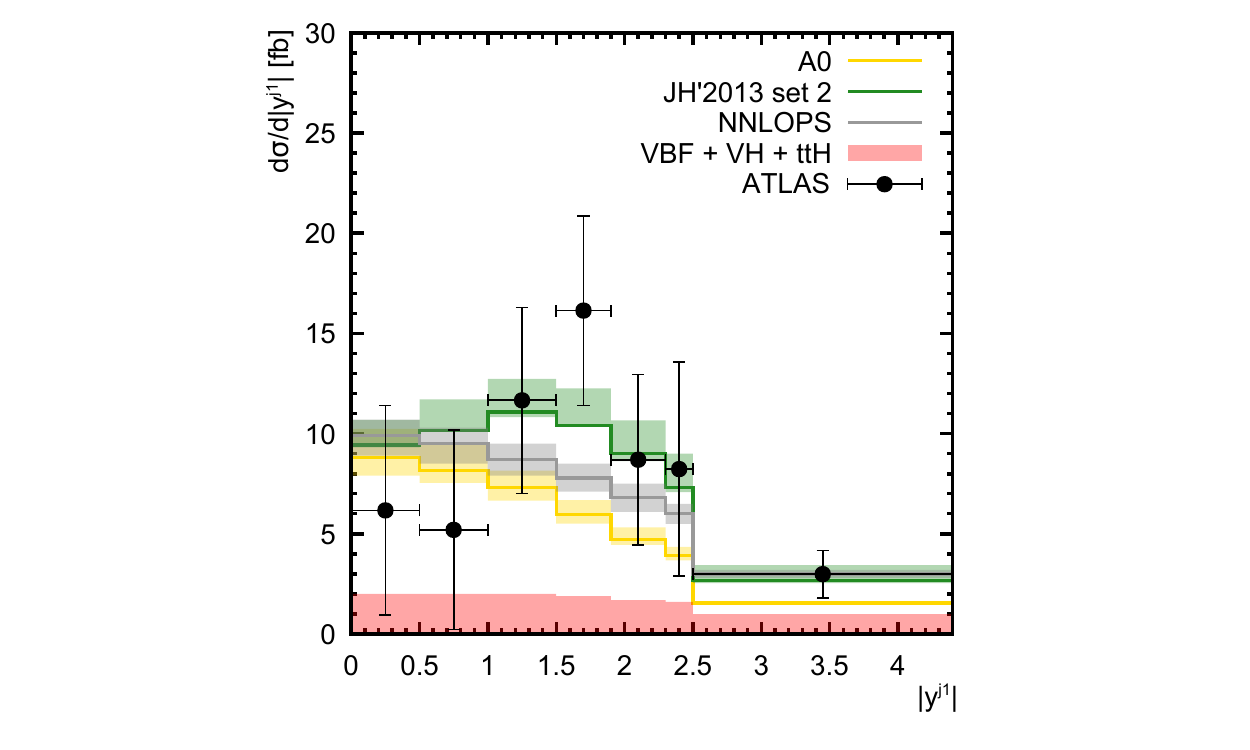}
\caption{The differential cross sections of associated Higgs boson and jet production 
(in the diphoton decay channel) at $\sqrt s = 13$~TeV as functions of $N_\text{jet}$,
leading jet transverse momentum and rapidity. 
The contributions from non-gluon fusion subprocesses and \textsc{nnlops} predictions are taken from\cite{1,5}. 
The experimental data are from CMS\cite{1} and ATLAS\cite{5}.}
\label{fig1}
\end{center}
\end{figure}

\begin{figure}
\begin{center}
\includegraphics[width=7.9cm]{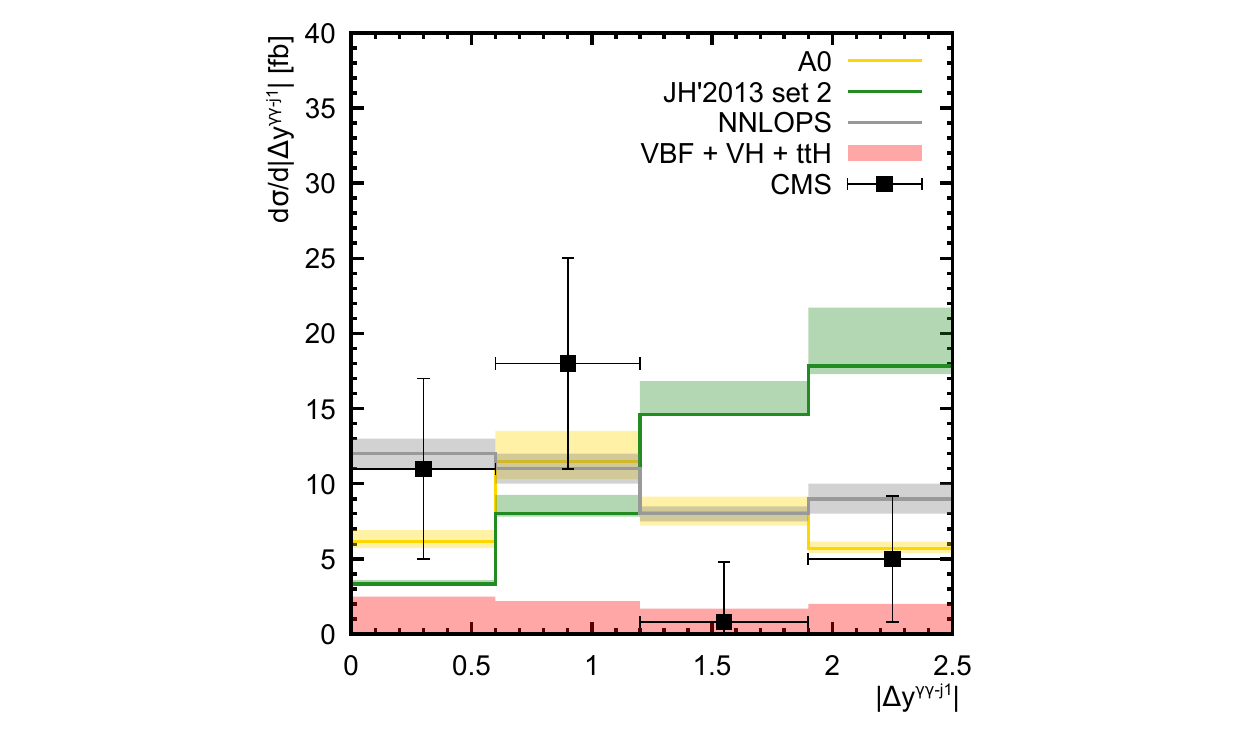}
\includegraphics[width=7.9cm]{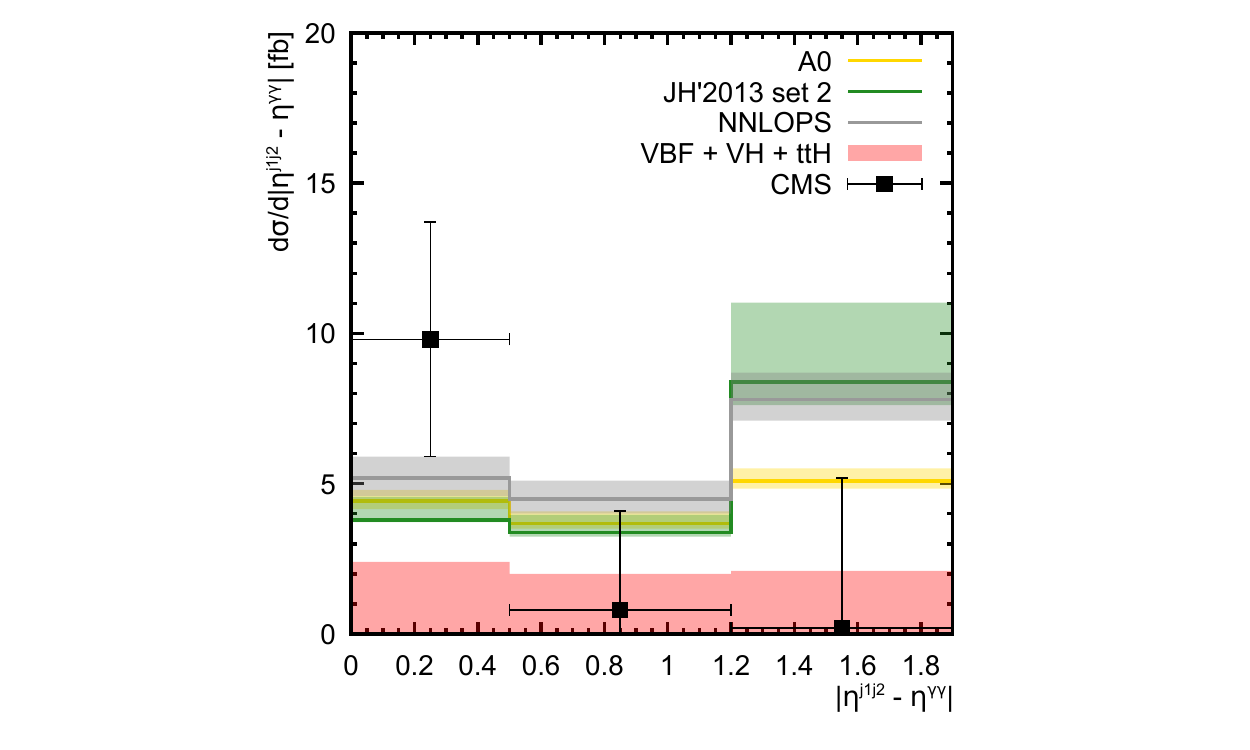}
\includegraphics[width=7.9cm]{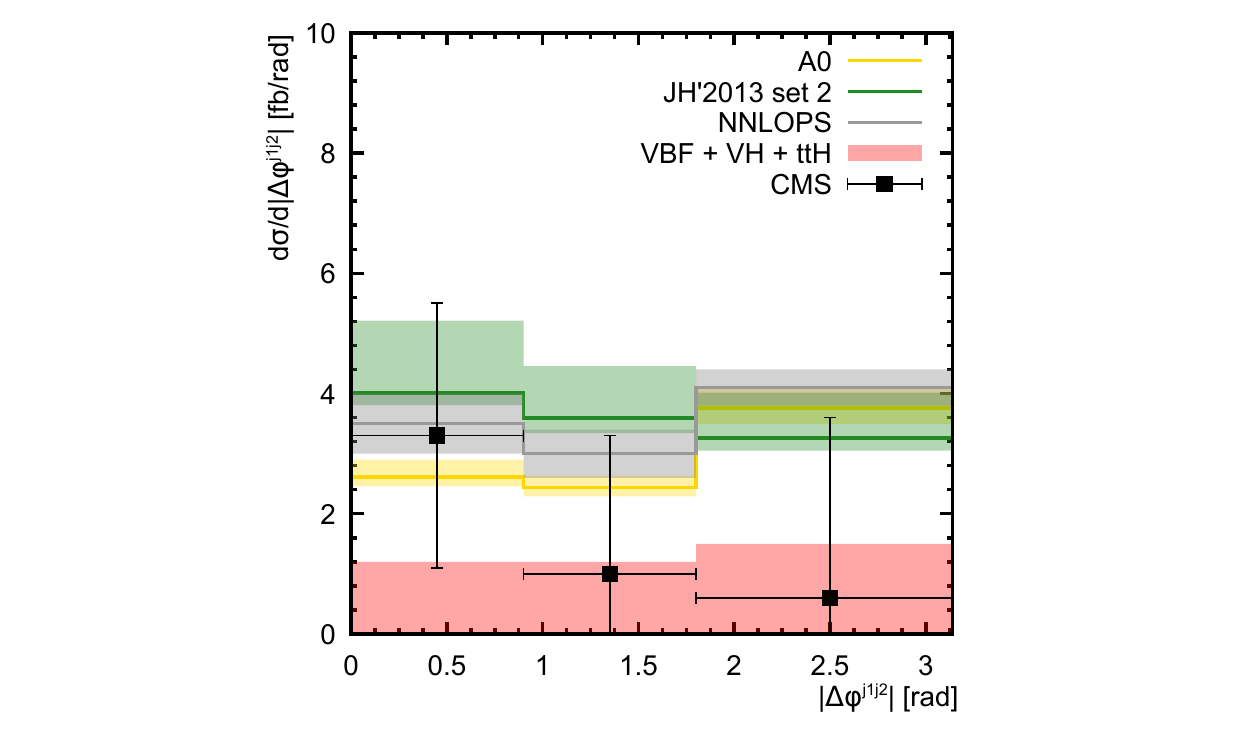}
\includegraphics[width=7.9cm]{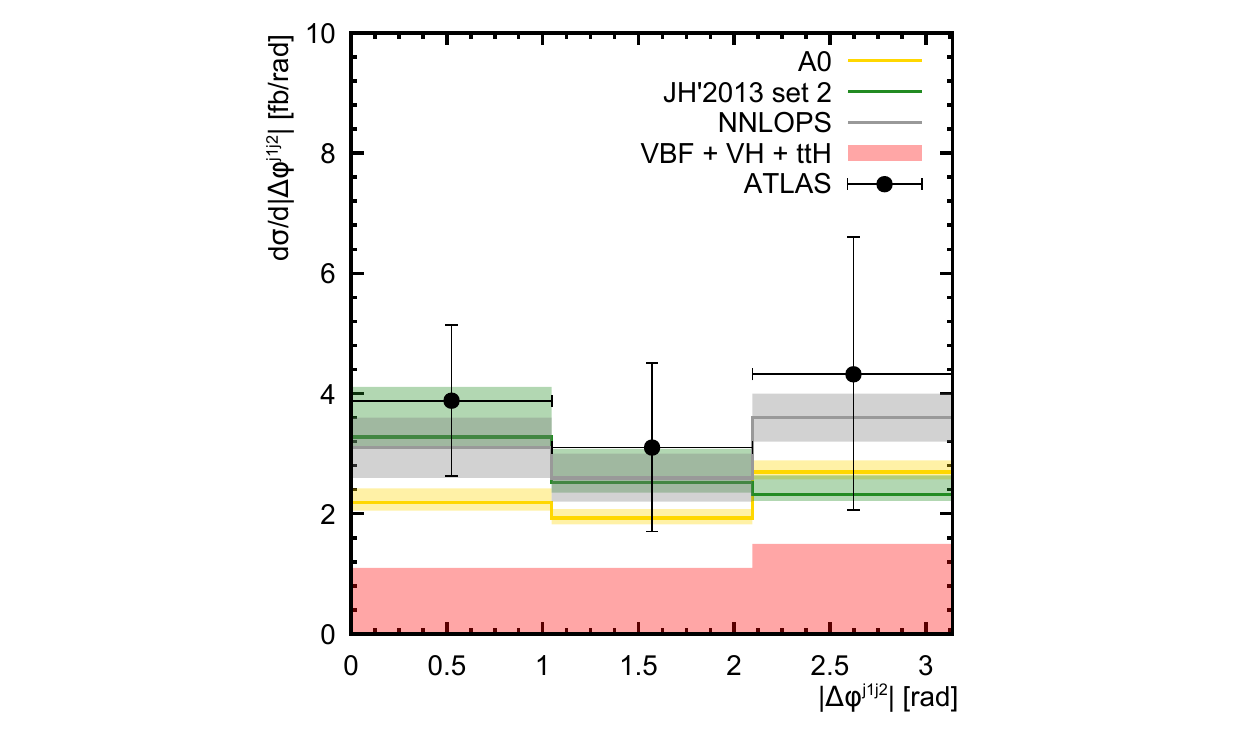}
\caption{The differential cross sections of associated Higgs boson 
and jet production (in the diphoton decay channel) at $\sqrt s = 13$~TeV
as functions of the rapidity difference between the diphoton system and 
leading jet $\Delta y^{\gamma\gamma j_1}$ and Zeppenfeld variable 
$|\eta^{j_1j_2} - \eta^{\gamma\gamma}|$.
The contributions from non-gluon fusion subprocesses and \textsc{nnlops} predictions are taken from\cite{1,5}. 
The experimental data are from CMS\cite{1} and ATLAS\cite{5}.}
\label{fig2}
\end{center}
\end{figure}

\begin{figure}
\begin{center}
\includegraphics[width=7.9cm]{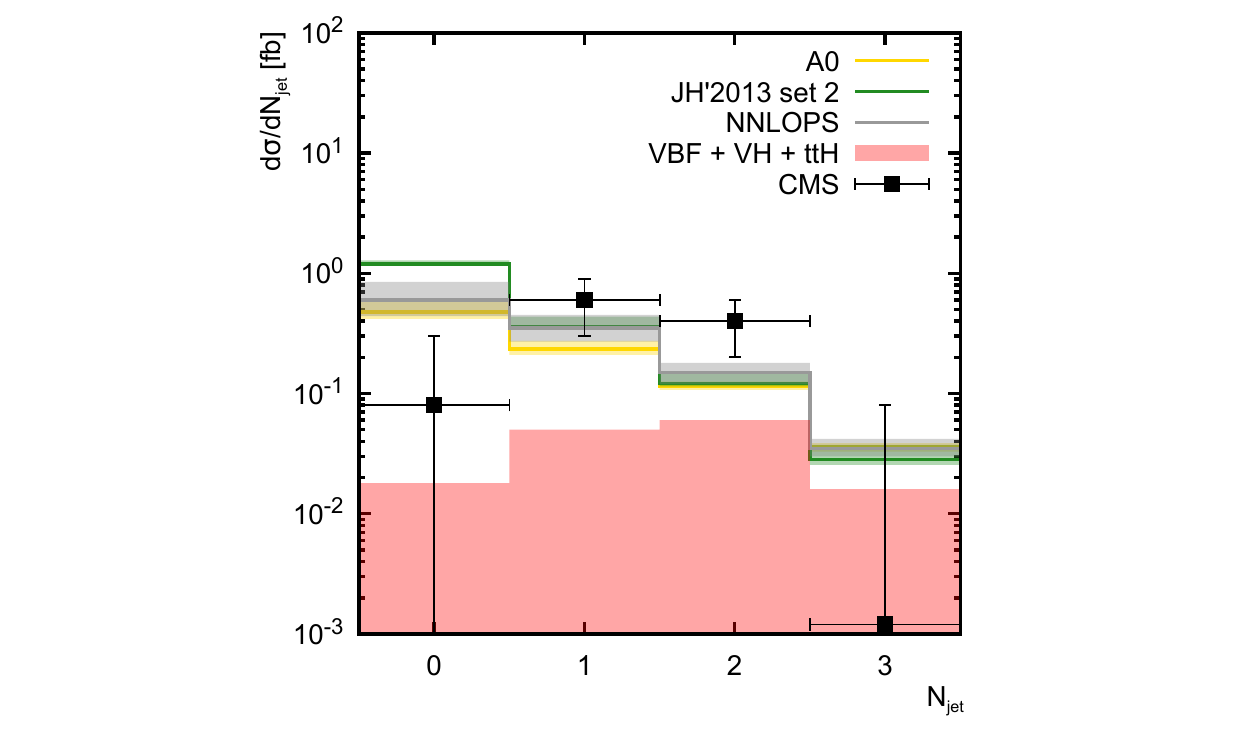}
\includegraphics[width=7.9cm]{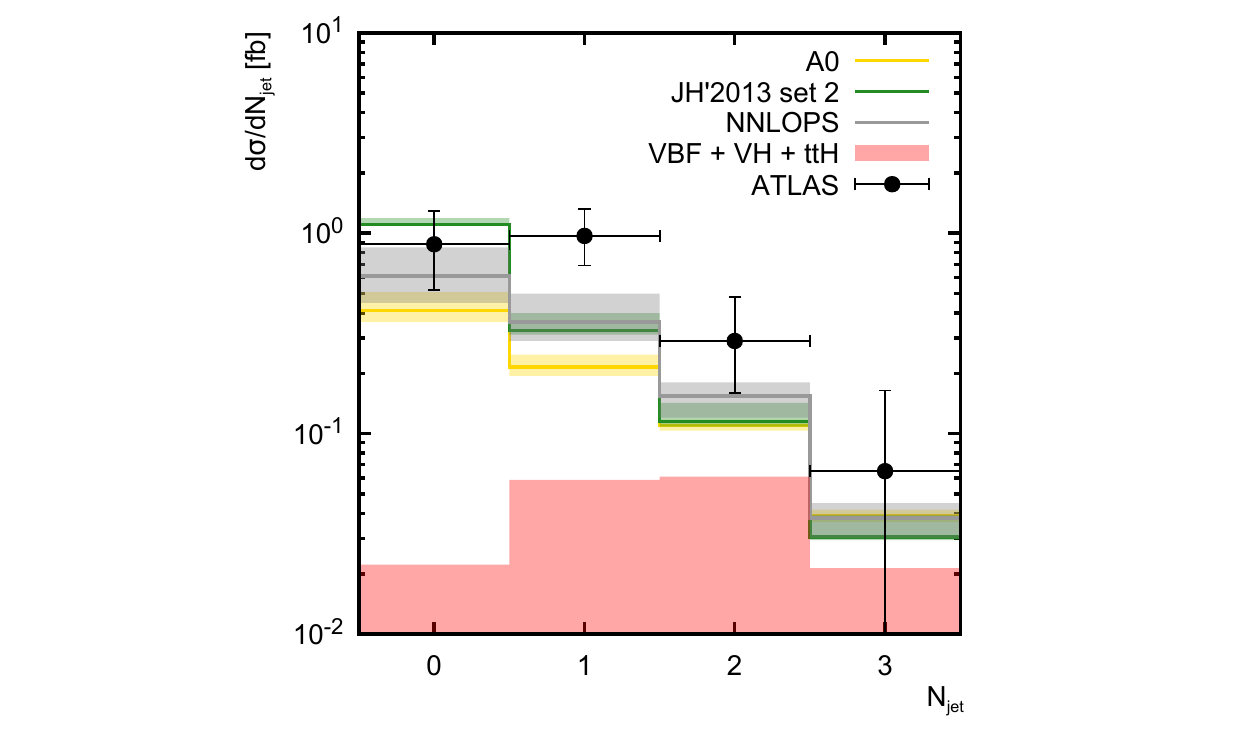}
\includegraphics[width=7.9cm]{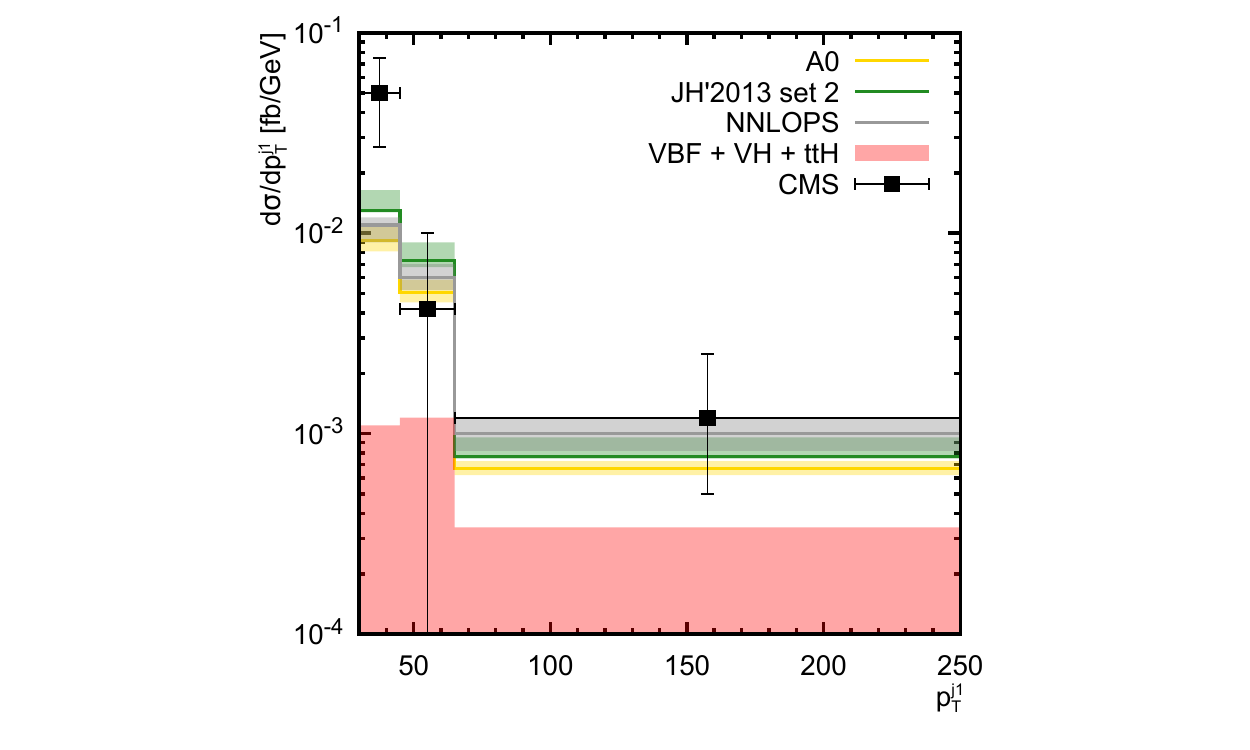}
\includegraphics[width=7.9cm]{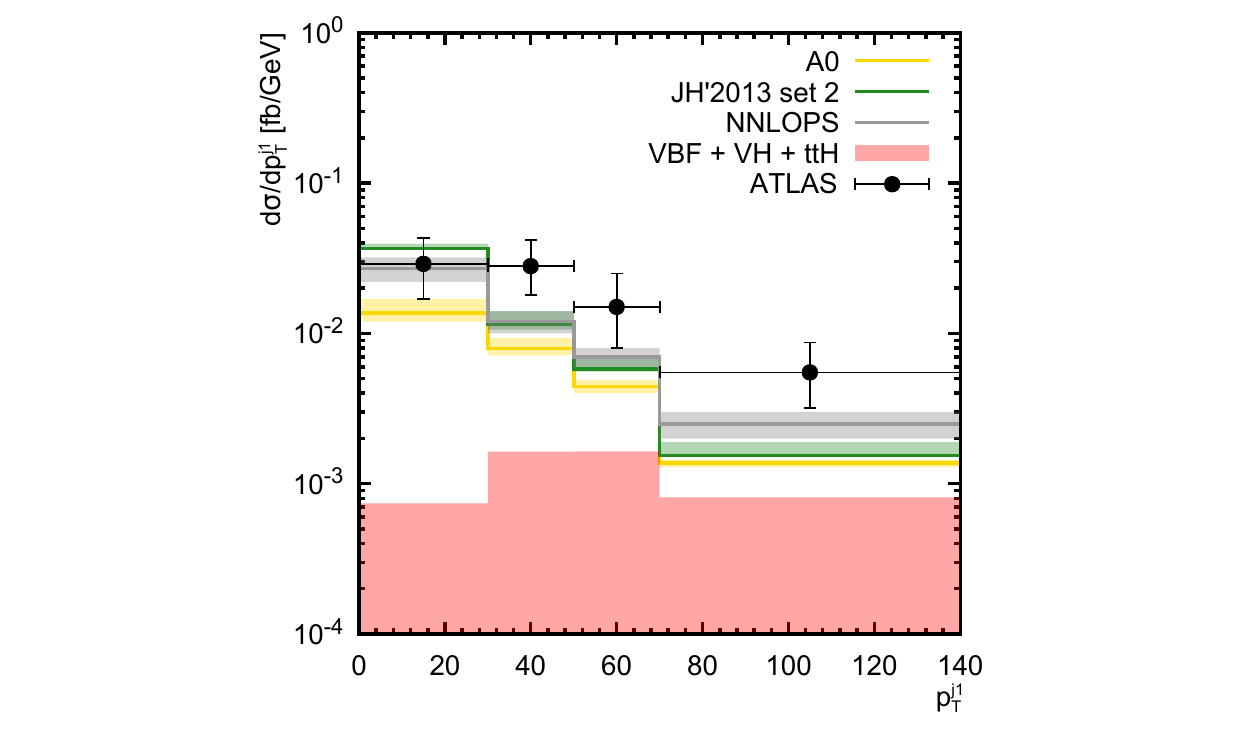}
\includegraphics[width=7.9cm]{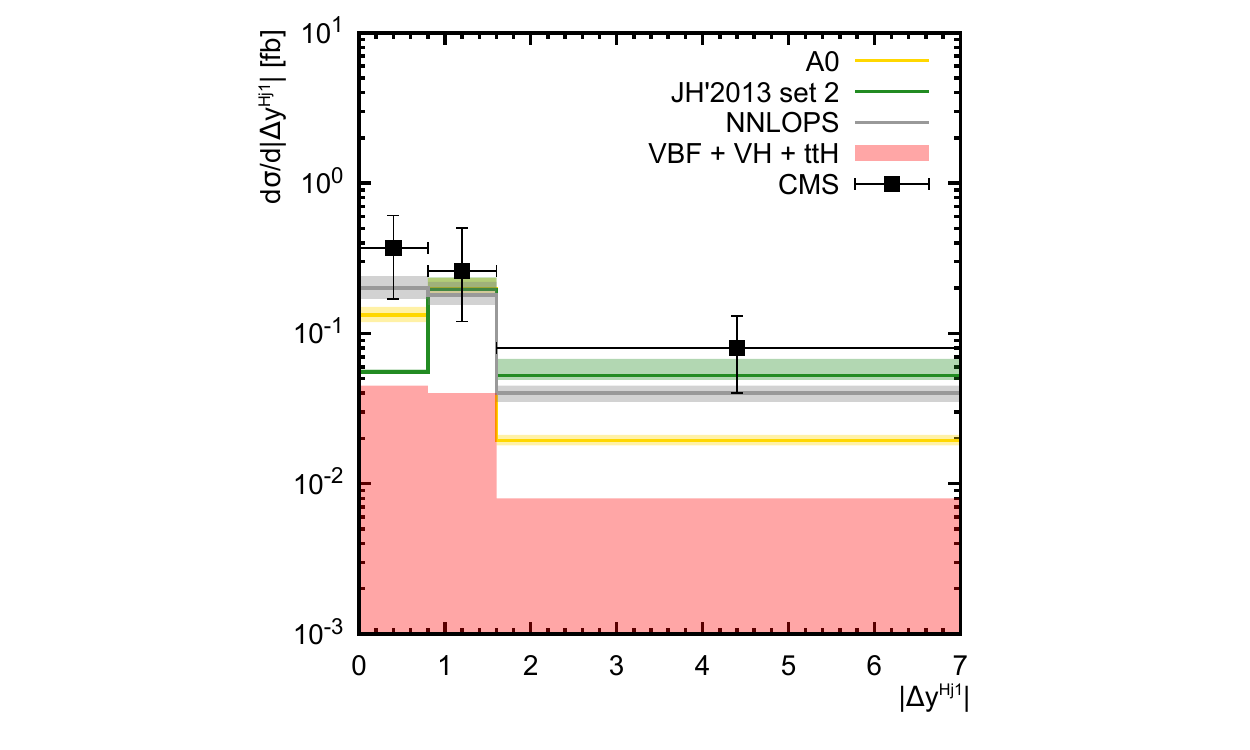}
\caption{The differential cross sections of associated Higgs boson and jet production (in the $H \to ZZ^*$ decay channel) at $\sqrt s = 8$~TeV
as functions of $N_\text{jet}$, leading jet transverse momentum and 
rapidity difference between the Higgs boson and leading jet $\Delta y^{Hj_1}$.
The contributions from non-gluon fusion subprocesses and \textsc{nnlops} predictions are taken from\cite{2,6}. 
The experimental data are from CMS\cite{2} and ATLAS\cite{6}.}
\label{fig3}
\end{center}
\end{figure}

\begin{figure}
\begin{center}
\includegraphics[width=7.9cm]{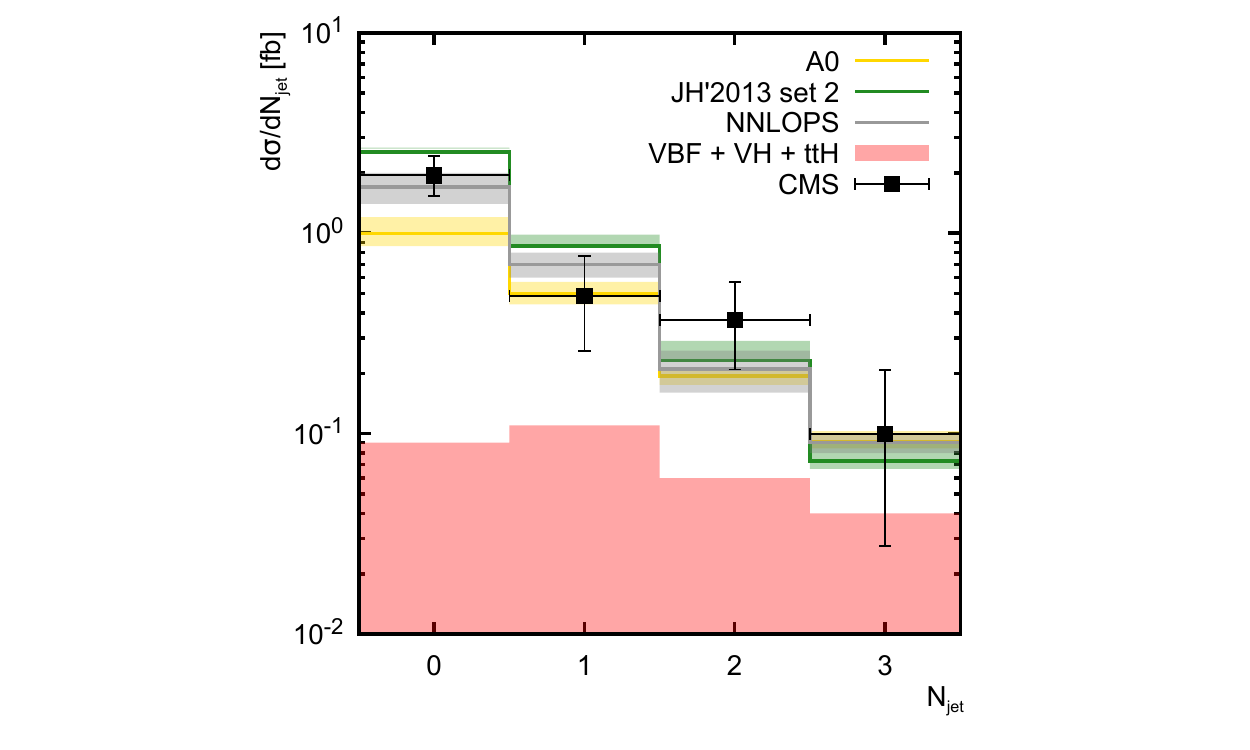}
\includegraphics[width=7.9cm]{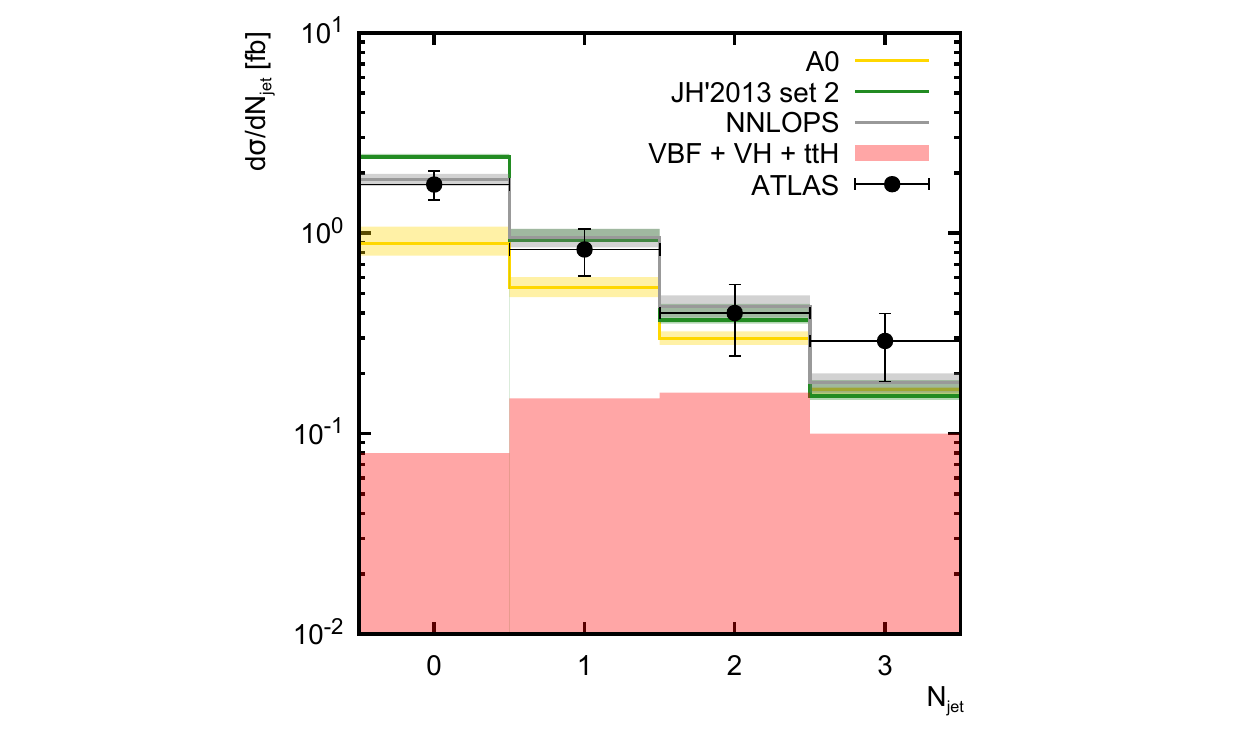}
\includegraphics[width=7.9cm]{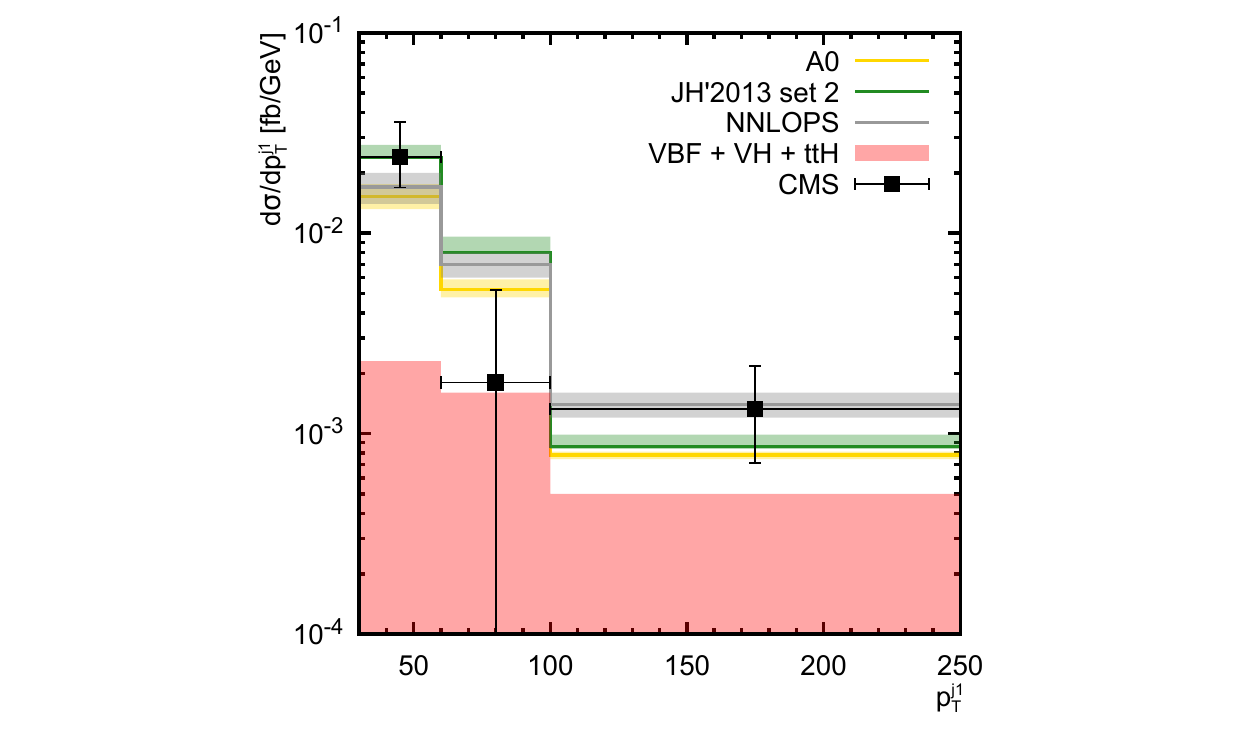}
\includegraphics[width=7.9cm]{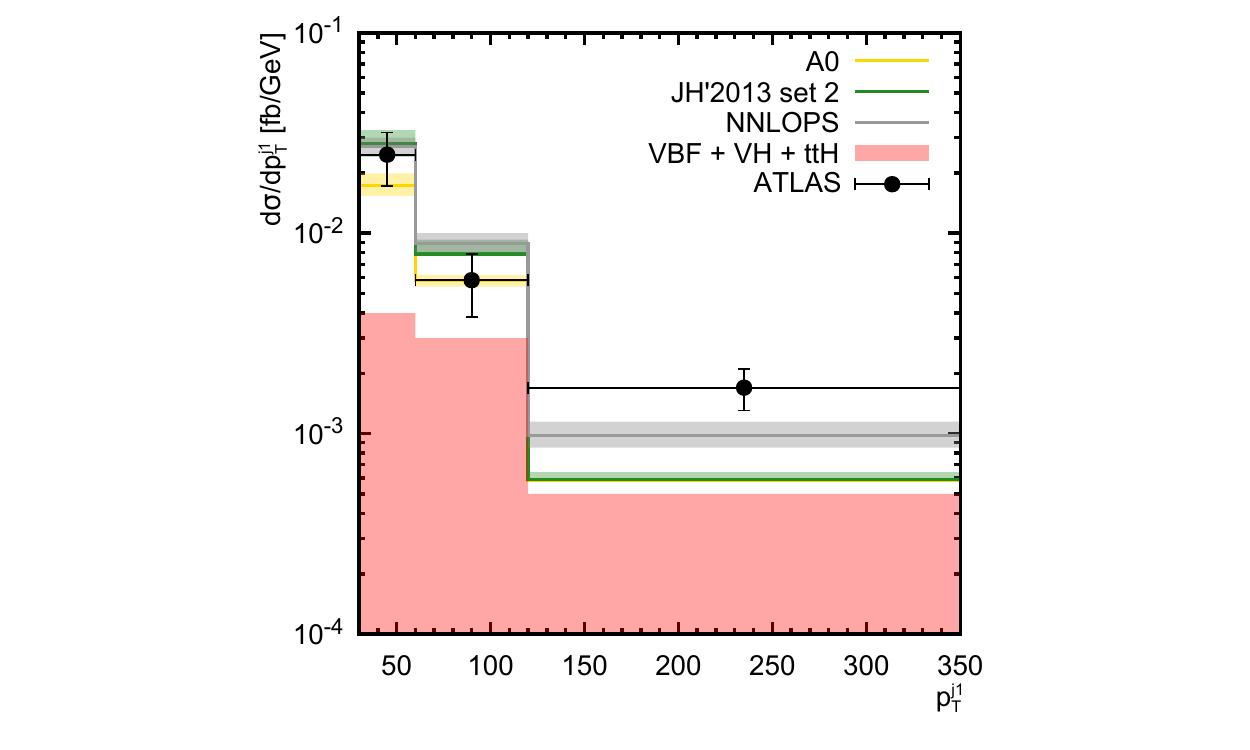}
\caption{The differential cross sections of associated Higgs boson and jet production (in the $H \to ZZ^*$ decay channel) at $\sqrt s = 13$~TeV
as functions of $N_\text{jet}$ and leading jet transverse momentum.
The contributions from non-gluon fusion subprocesses and \textsc{nnlops} predictions are taken from\cite{3,7}. 
The experimental data are from CMS\cite{3} and ATLAS\cite{7}.}
\label{fig4}
\end{center}
\end{figure}

\begin{figure}
\begin{center}
\includegraphics[width=7.9cm]{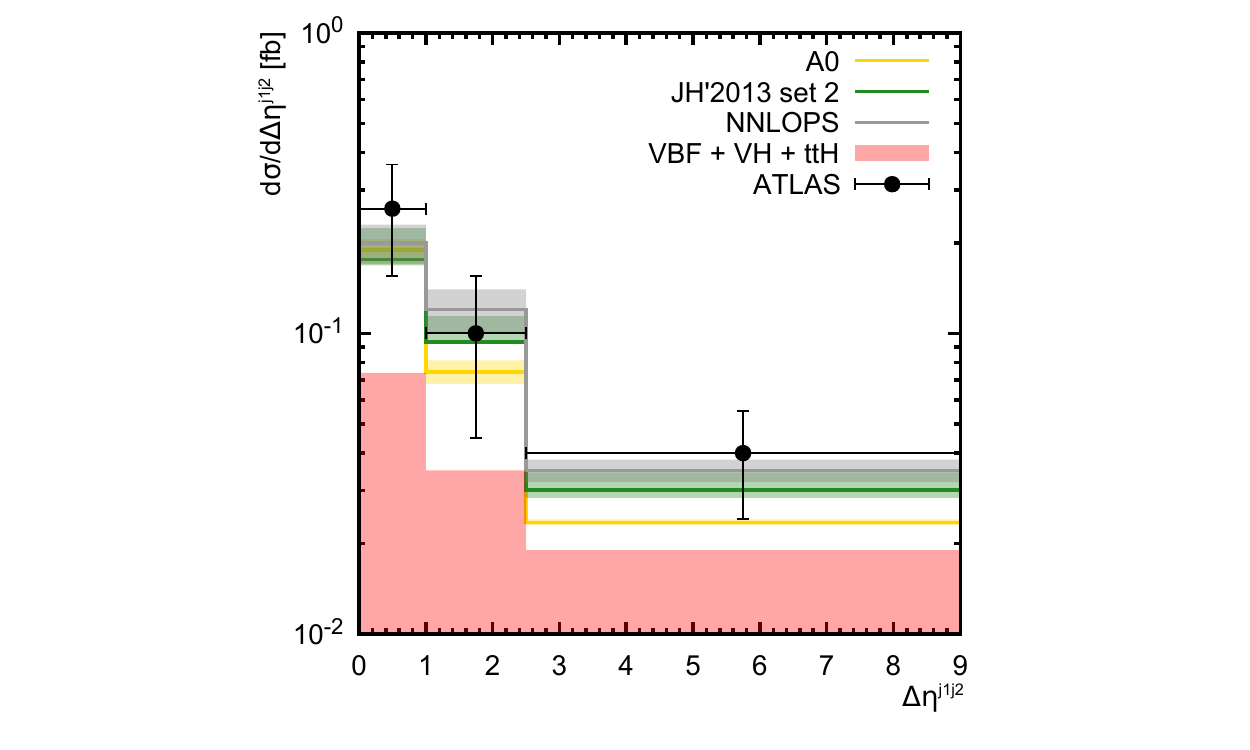}
\includegraphics[width=7.9cm]{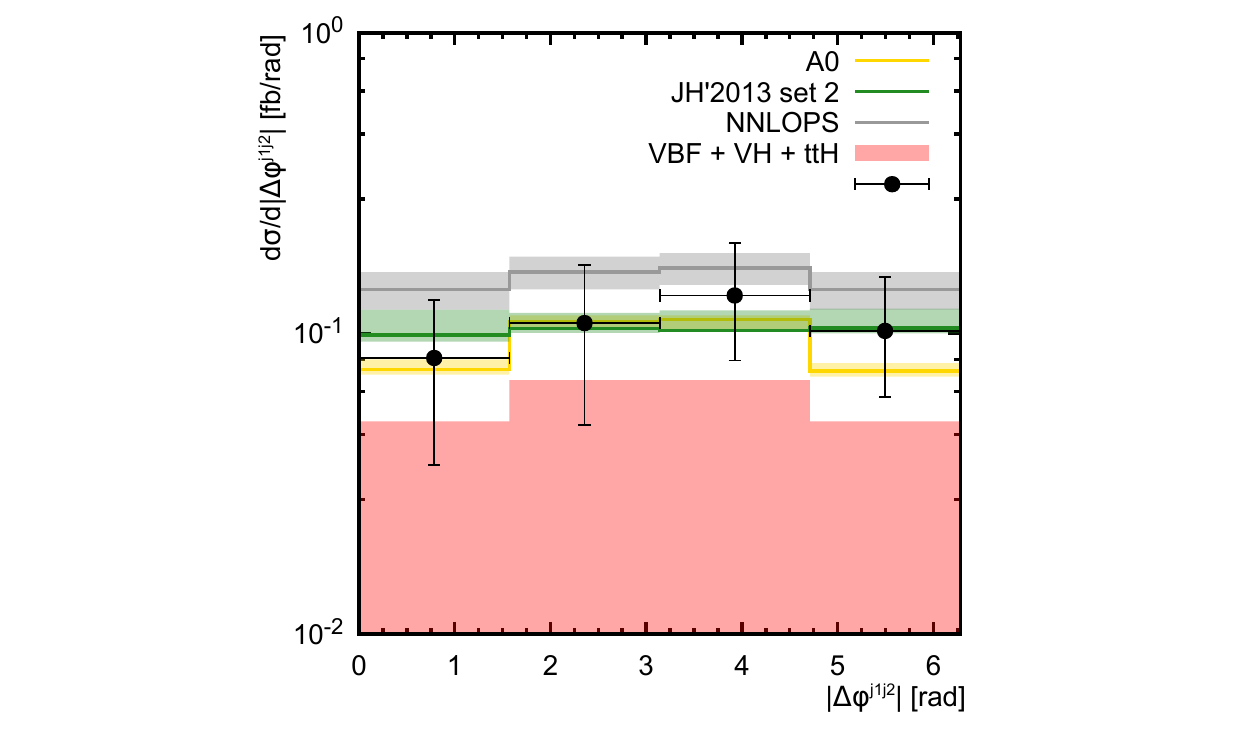}
\includegraphics[width=7.9cm]{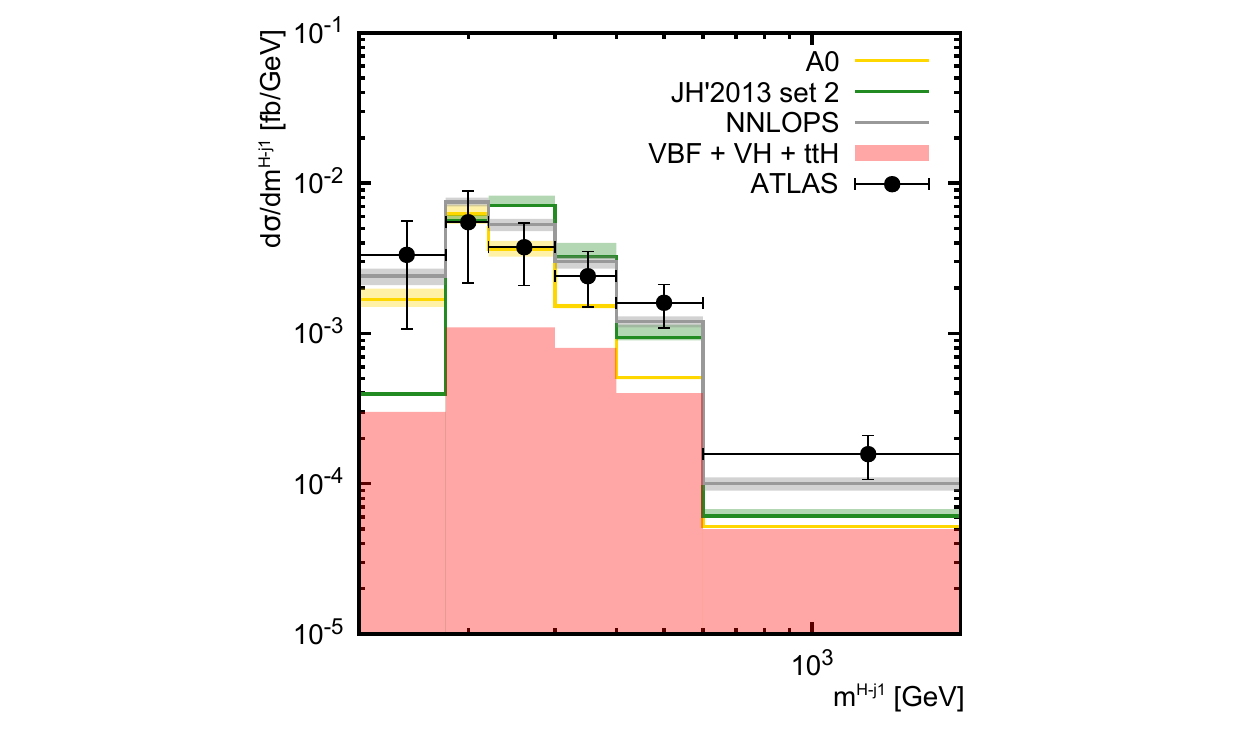}
\includegraphics[width=7.9cm]{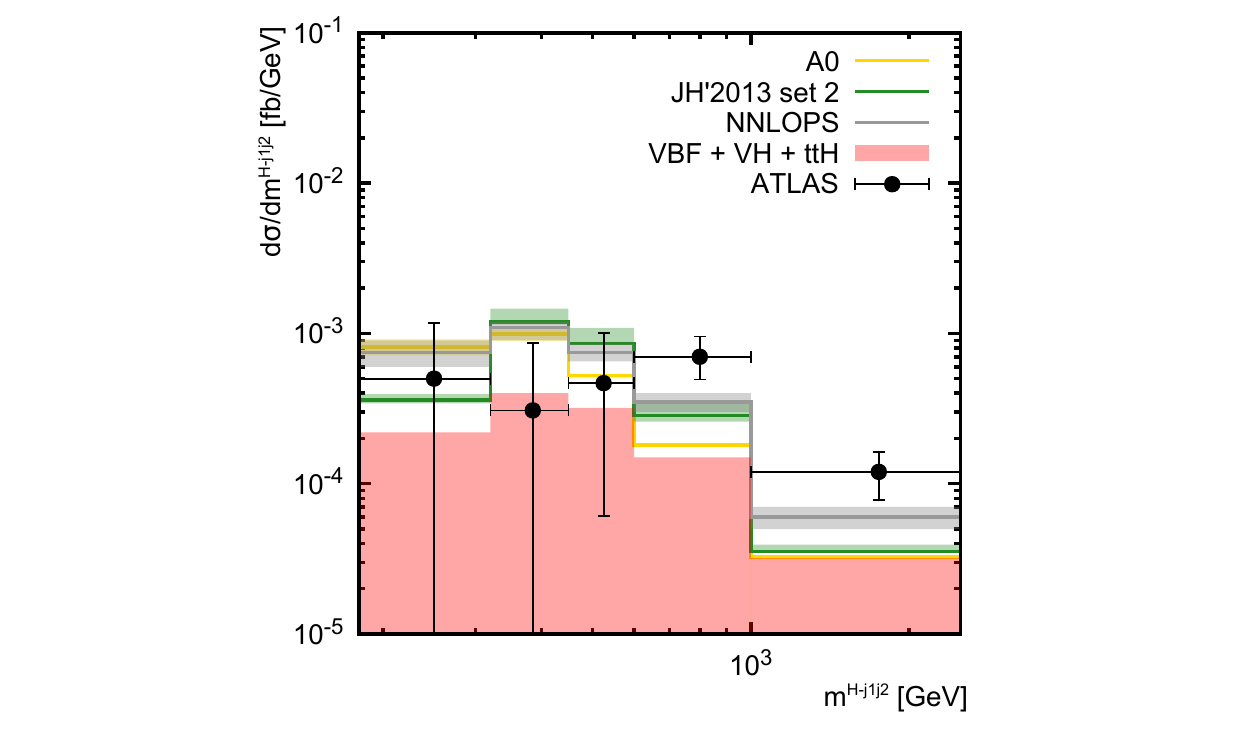}
\caption{The differential cross sections of associated Higgs boson and jet production (in the $H \to ZZ^*$ decay channel) at $\sqrt s = 13$~TeV
as functions of the rapidity and azimuthal angle difference between the leading and subleading jets, 
invariant masses of Higgs-leading jet system and Higgs-dijet system.
The contributions from non-gluon fusion subprocesses and \textsc{nnlops} predictions are taken from\cite{3,7}. 
The experimental data are from CMS\cite{3} and ATLAS\cite{7}.}
\label{fig5}
\end{center}
\end{figure}

\begin{figure}
\begin{center}
\includegraphics[width=7.9cm]{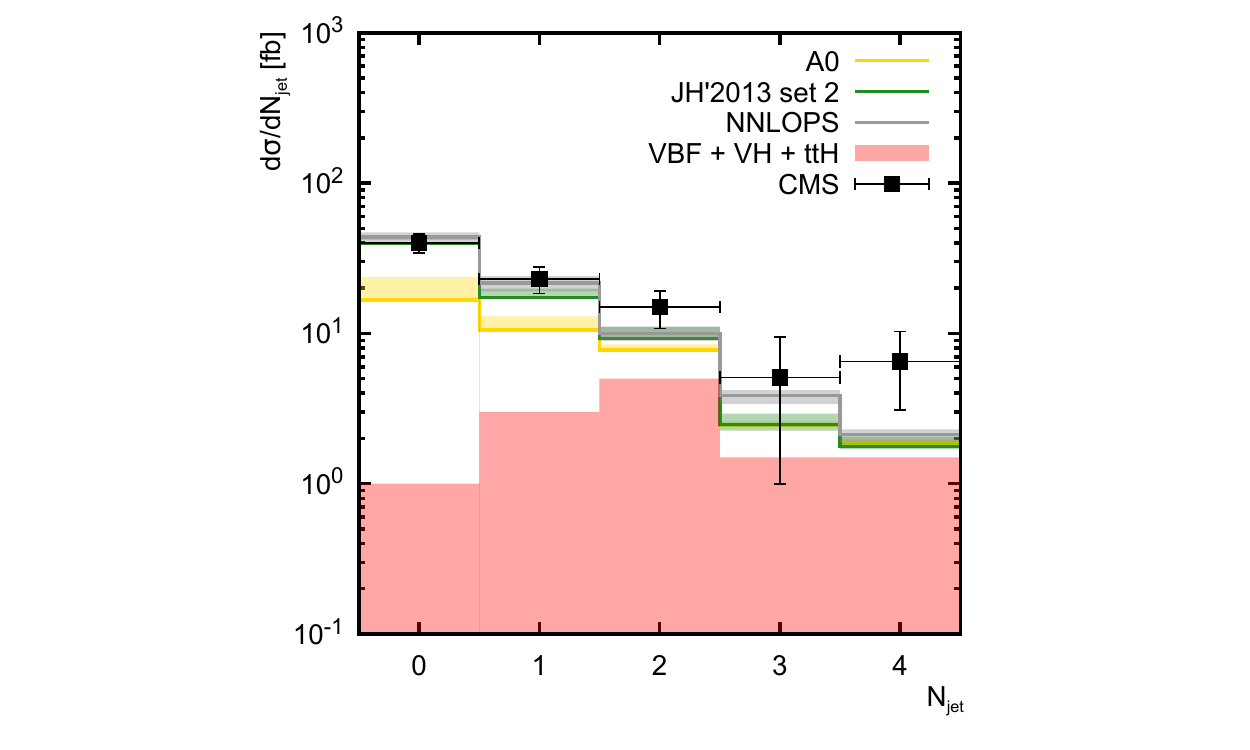}
\includegraphics[width=7.9cm]{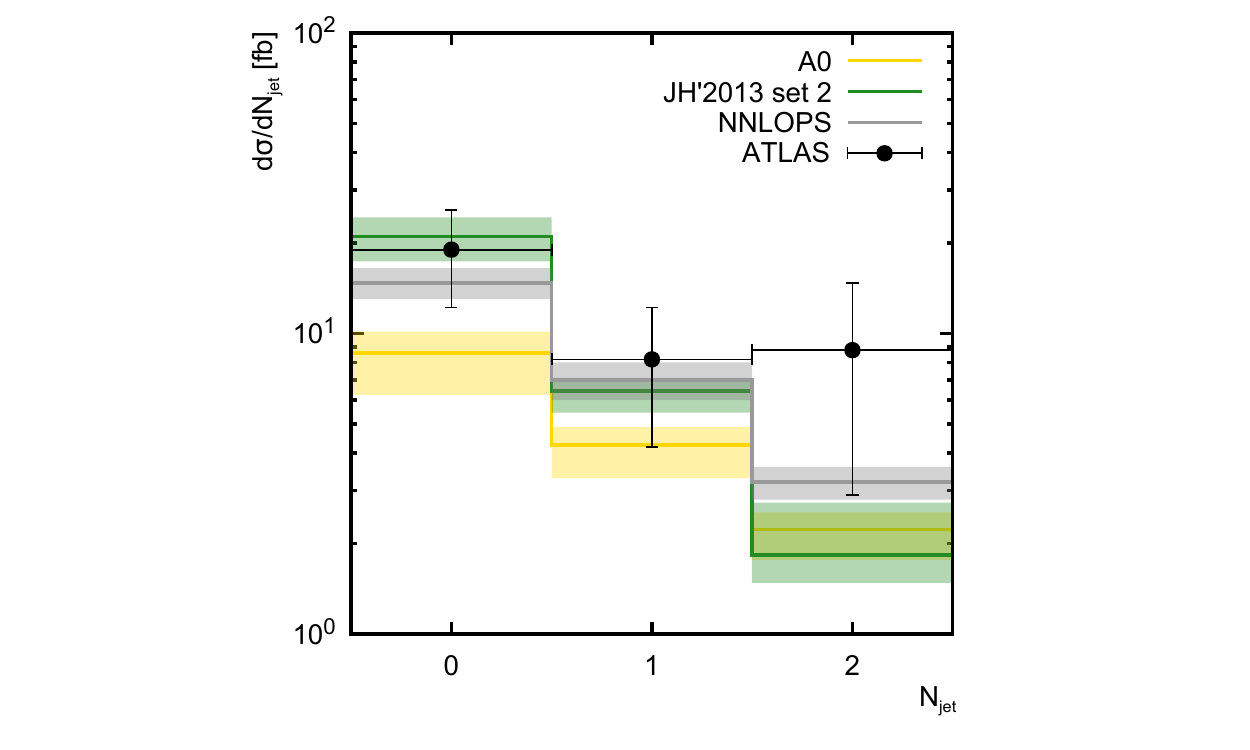}
\caption{The differential cross sections of associated Higgs boson and jet production (in the $H \to W^+W^-$ decay channel) at $\sqrt s = 8$ (right panel)
and $13$~TeV (left panel) as functions of $N_\text{jet}$.
The experimental data are from CMS\cite{4} and ATLAS\cite{8}.}
\label{fig6}
\end{center}
\end{figure}


\begin{thebibliography}{70}

\bibitem{1} CMS Collaboration, JHEP {\bf 1901}, 183 (2019). 
\bibitem{2} CMS Collaboration, JHEP {\bf 1604}, 005 (2016). 
\bibitem{3} CMS Collaboration, JHEP {\bf 1711}, 047 (2017). 
\bibitem{4} CMS Collaboration, arxiv:2007.01984 [hep-ex]. 
\bibitem{5} ATLAS Collaboration, Phys. Rev. D {\bf 98}, 052005 (2018). 
\bibitem{6} ATLAS Collaboration, Phys. Lett. B {\bf 738}, 234 (2014). 
\bibitem{7} ATLAS Collaboration, CERN-EP-2020-035; arXiv:2004.03969 [hep-ex]. 
\bibitem{8} ATLAS Collaboration, JHEP {\bf 1608}, 104 (2016). 
\bibitem{9} LHC Higgs Cross Section Working Group, arXiv:1610.07922 [hep-ph].
\bibitem{10} M.~Spira, A.~Djouadi, D.~Graudenz, P.M.~Zerwas, Nucl. Phys. B {\bf 453}, 17 (1995).
\bibitem{11} A.~Djouadi, M.~Spira, P.M.~Zerwas, Phys. Lett. B {\bf 264}, 440 (1991).
\bibitem{12} S.~Dawson, Nucl. Phys. B {\bf 359}, 283 (1991).
\bibitem{13} R.V.~Harlander, W.B.~Kilgore, Phys. Rev. Lett. {\bf 88}, 201801 (2002).
\bibitem{14} C.~Anastasiou, K.~Melnikov, Nucl. Phys. B {\bf 646}, 220 (2002).
\bibitem{15} V.~Ravindran, J.~Smith, W.L.~van Neerven, Nucl. Phys. B {\bf 665}, 325 (2003).
\bibitem{16} D.~de~Florian, G.~Ferrera, M.~Grazzini, D.~Tommasini,  JHEP {\bf 1206}, 132 (2012).
\bibitem{17} K.~Hamilton, P.~Nason, E.~Re, G.~Zanderighi, JHEP {\bf 1310}, 222 (2013).
\bibitem{18} C.~Anastasiou, C.~Duhr, F.~Dulat, E.~Furlan, T.~Gehrmann, F.~Herzog, A.~Lazopoulos, B.~Mistlberger, JHEP {\bf 1605}, 058 (2016).
\bibitem{19} U.~Aglietti, R.~Bonciani, G.~Degrassi, A.~Vicini, Phys. Lett. B {\bf 595}, 432 (2004).
\bibitem{20} S.~Actis, G.~Passarino, C.~Sturm, S.~Uccirati, Phys. Lett. B {\bf 670}, 12 (2008).
\bibitem{21} D.~de~Florian, M.~Grazzini, Phys. Lett. B {\bf 718}, 117 (2012).
\bibitem{22} C.~Anastasiou, S.~Buehler, F.~Herzog, A.~Lazopoulos, JHEP {\bf 1204}, 004 (2012).
\bibitem{23} J.~Baglio, A.~Djouadi, JHEP {\bf 1103}, 055 (2011).
\bibitem{24} S.~Catani, D.~de~Florian, M.~Grazzini, P.~Nason, JHEP {\bf 0307}, 028 (2003).
\bibitem{25} D.~de~Florian, G.~Ferrera, M.~Grazzini, D.~Tommasini, JHEP {\bf 1111}, 064 (2011).
\bibitem{26} S.~Catani, M.~Ciafaloni, F.~Hautmann, Nucl. Phys. B {\bf 366}, 135 (1991);\\
  J.C.~Collins, R.K.~Ellis, Nucl. Phys. B {\bf 360}, 3 (1991).
\bibitem{27} L.V.~Gribov, E.M.~Levin, M.G.~Ryskin, Phys. Rep. {\bf 100}, 1 (1983);\\
  E.M.~Levin, M.G.~Ryskin, Yu.M.~Shabelsky, A.G.~Shuvaev, Sov. J. Nucl. Phys. {\bf 53}, 657 (1991).
\bibitem{28} E.A.~Kuraev, L.N.~Lipatov, V.S.~Fadin, Sov. Phys. JETP {\bf 44}, 443 (1976);\\
  E.A.~Kuraev, L.N.~Lipatov, V.S.~Fadin, Sov. Phys. JETP {\bf 45}, 199 (1977);\\
  I.I.~Balitsky, L.N.~Lipatov, Sov. J. Nucl. Phys. {\bf 28}, 822 (1978).
\bibitem{29} M.~Ciafaloni, Nucl. Phys. B {\bf 296}, 49 (1988);\\
  S.~Catani, F.~Fiorani, G.~Marchesini, Phys. Lett. B {\bf 234}, 339 (1990);\\
  S.~Catani, F.~Fiorani, G.~Marchesini, Nucl. Phys. B {\bf 336}, 18 (1990);\\
  G.~Marchesini, Nucl. Phys. B {\bf 445}, 49 (1995).
\bibitem{30} R.~Angeles-Martinez et al., Acta Phys. Polon. B {\bf 46}, 2501 (2015).
\bibitem{31} Small-$x$ Collaboration, Eur. Phys. J. C {\bf 25}, 77 (2002);\\
  Small-$x$ Collaboration, Eur. Phys. J. C {\bf 35}, 67 (2004);\\
  Small-$x$ Collaboration, Eur. Phys. J. C {\bf 48}, 53 (2006).
\bibitem{32} A.~Gawron, J.~Kwiecinski, Phys. Rev. D {\bf 70}, 014003 (2004).
\bibitem{33} A.V.~Lipatov, N.P.~Zotov, Eur. Phys. J. C {\bf 44}, 559 (2005).
\bibitem{34} R.S.~Pasechnik, O.V.~Teryaev, A.~Szczurek, Eur. Phys. J. C {\bf 47}, 429 (2006).
\bibitem{35} G.~Watt, A.D.~Martin, M.G.~Ryskin, Phys. Rev. D {\bf 70}, 014012 (2004).
\bibitem{36} F.~Hautmann, Phys. Lett. B {\bf 535}, 159 (2002).
\bibitem{37} H.~Jung, Mod. Phys. Lett. A {\bf 19}, 1 (2004).
\bibitem{38} A.V.~Lipatov, M.A.~Malyshev, N.P.~Zotov, Phys. Lett. B {\bf 735}, 79 (2014).
\bibitem{39} R.~Islam, M.~Kumar, V.S.~Rawoot, Eur. Phys. J. C {\bf 79}, 181 (2019).
\bibitem{40} A.~Szczurek, M.~Luszczak, R.~Maciula, Phys. Rev. D {\bf 90}, 094023 (2014).
\bibitem{41} N.A.~Abdulov, A.V.~Lipatov, M.A.~Malyshev, Phys. Rev. D {\bf 97}, 054017 (2018).
\bibitem{42} N.A.~Abdulov, H.~Jung, A.V.~Lipatov, G.I.~Lykasov, M.A.~Malyshev, Phys. Rev. D {\bf 98}, 054010 (2018).
\bibitem{43} A.V.~Kotikov, A.V.~Lipatov, B.G.~Shaikhatdenov, P.~Zhang, JHEP {\bf 02}, 028 (2020).
\bibitem{44} S.~Dooling, F.~Hautmann, H.~Jung, Phys. Lett. B {\bf 736}, 293 (2014).
\bibitem{45} H.~Jung, S.P.~Baranov, M.~Deak, A.~Grebenyuk, F.~Hautmann, M.~Hentschinski, A.~Knutsson,  
  M.~Kramer, K.~Kutak, A.V.~Lipatov, N.P.~Zotov, Eur. Phys. J. C {\bf 70}, 1237 (2010).
\bibitem{46} J.R.~Ellis, M.K.~Gaillard, D.V.~Nanopoulos, Nucl. Phys. B {\bf 106}, 292 (1976).
\bibitem{47} M.A.~Shifman, A.I.~Vainstein, M.B.~Voloshin, V.I.~Zakharov, Sov. J. Nucl. Phys. {\bf 30}, 711 (1979).
\bibitem{48} A.V.~Lipatov, M.A.~Malyshev, S.P.~Baranov, Eur. Phys. J. C {\bf 80}, 330 (2020); \\
  https://theory.sinp.msu.ru/dokuwiki/doku.php/pegasus/news
\bibitem{49} F.~Hautmann, H.~Jung, Nucl. Phys. B {\bf 883}, 1 (2014).
\bibitem{50} H.~Jung, arXiv:hep-ph/0411287.
\bibitem{51} http://tmd.hepforge.org
\bibitem{52} M.~Cacciari, G.P.~Salam, G.~Soyez, Eur. Phys. J. C {\bf 72}, 1896 (2012).
\bibitem{53} J.~Alwall, A.~Ballestrero, P.~Bartalini, S.~Belov, E.~Boos, A.~Buckley, J.M.~Butterworth, 
  L.~Dudko, S.~Frixione, L.~Garren, S.~Gieseke, A.~Gusev, I.~Hinchliffe, J.~Huston, B.~Kersevan, 
  F.~Krauss, N.~Lavesson, L.~L\"onnblad, E.~Maina, F.~Maltoni, M.L.~Mangano, F.~Moortgat, S.~Mrenna, 
  C.G.~Papadopoulos, R.~Pittau, P.~Richardson, M.H.~Seymour, A.~Sherstnev, T.~Sj\"ostrand, P.~Skands, 
  S.R.~Slabospitsky, Z.~Wcas, B.R.~Webber, M.~Worek, D.~Zeppenfeld, Comput. Phys. Commun. {\bf 176}, 300 (2007).
\bibitem{54} PDG Collaboration, Phys. Rev. D {\bf 98}, 030001 (2018).
\bibitem{55} D.L.~Rainwater, R.~Szalapski, D.~Zeppenfeld, Phys. Rev. D {\bf 54}, 6680 (1996).
  
\end{thebibliography}
\end{document}